\documentclass[a4paper,12pt]{article}
\usepackage[english]{babel}
\usepackage[intlimits]{amsmath}
\usepackage{cite}
\usepackage{epsfig}
\usepackage{epsf}

\newcommand{\be}{\begin{equation}}
\newcommand{\ee}{\end{equation}}
\newcommand{\ben}{\begin{equation*}}
\newcommand{\een}{\end{equation*}}

\newcommand{\Li}{\mathrm{Li}}

\renewcommand{\not}[1]{#1\mspace{-8mu}\slash}

\begin{document}
\begin{titlepage}
\hskip 11cm
\begin{center}
{\bf Effective vertex of quark production in collision of Reggeized quark and
gluon$^{~\ast}$ }
\end{center}

\vskip 0.5cm \centerline{ M.~G.~Kozlov$^{\dag}$
 and A.~V.~Reznichenko$^{\ddag}$} \vskip
.3cm
\begin{center}
{\sl Budker Institute of  Nuclear Physics of Siberian Branch Russian Academy of
Sciences, Novosibirsk, 630090 Russia,\\ Novosibirsk State University, Novosibirsk,
630090 Russia}
\end{center}
\vskip 1cm

\begin{abstract}
We calculated the effective vertex of the quark production in the collision of
Reggeized quark and Reggeized gluon in the next-to-leading order (NLO). The vertex
in question is the missing component of the multi-Regge NLO amplitudes with the
quark and gluon exchanges in $t_i$ channels. The calculation allows us to develop
the bootstrap approach to the quark Reggeization proof within next-to-leading
logarithmic approximation.

\end{abstract}

\hrule \vskip .3cm \noindent $^{\ast}${\it  Work is supported by the Russian
Scientific Foundation,} (grant RFBR  13-02-01023, 15-02-07893) and by the Dynasty Foundation.

\vfill $
\begin{array}{ll} ^{\dag}\mbox{{\it e-mail address:}} &
\mbox{M.G.Kozlov@inp.nsk.su}\\
^{\ddag}\mbox{{\it e-mail address:}} &
\mbox{A.V.Reznichenko@inp.nsk.su}\\
\end{array}
$
\end{titlepage}

\section{Introduction}

As it is well known  the multi-Regge form of amplitudes at high energies is a base
of various theoretical constructions in quantum chromodynamics (QCD) and
supersymmetric Yang-Mills theories (SYM). The most famous application of the form
resulted in the Balitsky-Fadin-Kuraev-Lipatov (BFKL)
\cite{Balitsky:1978ic,Fadin:1975cb, Kuraev:1976ge, Kuraev:1977fs} approach to the
semi-hard processes description in QCD. The simplicity of this form has recently
given  the powerful tool for various factorization formulae verification in SYM.

Let us now remind the state of art for the Reggeization hypothesis proof. In QCD
gluon Reggeization hypothesis (i.e. the multi-Regge form with only gluon exchanges
in all $t_i$ channels) was proved in leading logarithmic approximation (LLA) by the
authors of the BFKL approach roughly  forty years ago. The analyticity and t-channel
unitarity were the principal tools of this proof \cite{Lipatov:1976zz}. It proved to
be the strong base of the BFKL approach in the leading approximation.

In next-to-leading logarithmic approximation (NLA) we  developed the general method
based on the compatibility of the hypothetical multi-Regge form of the amplitude
with s-channel unitarity  \cite{Fadin:2006bj}. The compatibility is formulated as a
series of so-called bootstrap relations which fulfillment ensures validity of the
multi-Regge form order by order. The same method turned out to be fruitful in the
proof  of the multi-Regge form with the quark exchanges in LLA \cite{BF:2006}. Then
we used the bootstrap approach to prove the NLA gluon Reggeization hypothesis in
QCD. The calculation of the quark  and gluon  one-loop corrections to all bootstrap
components gave us possibility to verify all bootstrap relations
\cite{Fadin:2003xs,FKR:2010-1,FKR:2010-2,FKR:2010-3}. Once again our general method
was successfully  applied  to prove  NLA gluon Reggeization within supersymmetric
Yang-Mills (SYM) theories with arbitrary ${\cal N}$ and in the theories with general
form of Yukawa interaction \cite{FKR:2013}.  Theoretically our bootstrap approach is
applicable for the quark Reggeization NLA proof. But the only unknown component of
the NLA amplitude is the Reggeon(G)-Reggeon(Q)-quark one-loop vertex $\gamma_{{\cal
G}_1{\cal Q}_2}^{Q}$.

The only missing link of the recurrent bootstrap procedure is the ``initial
condition''. For NLA it is  one-loop amplitude with arbitrary leg number $n$. We
supposed  these amplitudes to have the correct factorized form corresponding to the
multi-Regge anzats. It has been verified for small $n$ and  should be proved in
general.

The main goal  of our investigation is  to complete the quark Reggeization
hypothesis formulation in NLA. For this purpose one should know all effective
Reggeon vertices appearing in the multi-Regge form with the quark exchanges up to
next-to-leading order.  The vertex $\gamma_{{\cal G}_1{\cal Q}_2}^{Q}$ in NLO is the
final uncalculated component of the amplitude. Another aspect of interest is the
construction of the evolution equation kernel in NLO for the Reggeized quark. The
vertex $\gamma_{{\cal G}_1{\cal Q}_2}^{Q}$ is the final ingredient for the kernel
construction. The kernel is of concern since its conformal properties in SYM theory
can illuminate the integrability property of this theory and give the connection
between different approaches as it occurred  for the gluon NLO kernel. Finally,
there are some processes (as for instance, recharging processes  $p+p \rightarrow n
+ \Delta^{++}$, $p+\bar{p} \rightarrow n + \bar{n} $ with u- and d- quark exchanges)
where the quark exchange amplitude (subleading in comparison with the gluon one)
might dominantly contribute. Our Reggeon vertices are of some phenomenological
interest for these processes.

The article is organized as follows. The second section is devoted to our method explanation and the kinematics description. The third section presents the Reggeization hypothesis, components of our Regge amplitude, and Lorentz and color structures of our one-loop amplitude in the multi-Regge kinematics. In the next section we present the result of our calculation both in fragmentation form reproducing different components of our Regge amplitude and as the full expression. At the end of fourth section we present the resulting expression for the required vertex $\gamma_{{\cal G}_1{\cal Q}_2}^{Q}$. In the Appendix we introduce the technique of the loop integration and give the explicit expressions for the master-intergals of our calculation. 

\section{Amplitude of the quark production in MRK}
\label{section-Amplitude}
\begin{figure}[t]
\begin{center}%
\epsfig{file=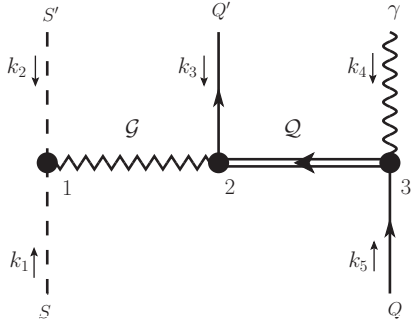,width=6cm} \caption{Regge amplitude of the process
$SQ\rightarrow S'Q'\gamma$. ${\cal G}$ and ${\cal Q}$ are Reggeized gluon in
$t_1$-channel and Reggeized quark in $t_2$-channel respectively. Here blob 1 is the
effective vertex $\Gamma_{S'S}^{{\cal G}}$, blob 2~--- unknown vertex $\gamma_{{\cal
G} {\cal Q}}^{Q'}$, and blob 3~--- vertex $\Gamma_{\gamma' Q}^{{\cal Q}}$.}
\label{fig:1}
\end{center}
\end{figure}

There are several stages  in NLO effective vertex $\gamma_{{\cal G}_1{\cal
Q}_2}^{Q}$ finding in the next-to-leading order. To calculate this Reggeon vertex we
can consider any simple process with this vertex in one-loop approximation. We
choose the amplitude $SQ\rightarrow S'Q'\gamma$ of the scalar, quark, and photon
production in scalar and quark collision: see Fig.~ \ref{fig:1}. It does not matter
for the vertex calculation whether we analyze amplitude in Yang-Mills theory with
$N_c$ gluons, the photon, $n_f$ quarks (in the fundamental color representation),
and $n_s$ scalars (in the adjoint representation) or simply QCD amplitude.

\begin{figure}
\begin{center}
\begin{minipage}[t]{34mm}
\epsfig{file=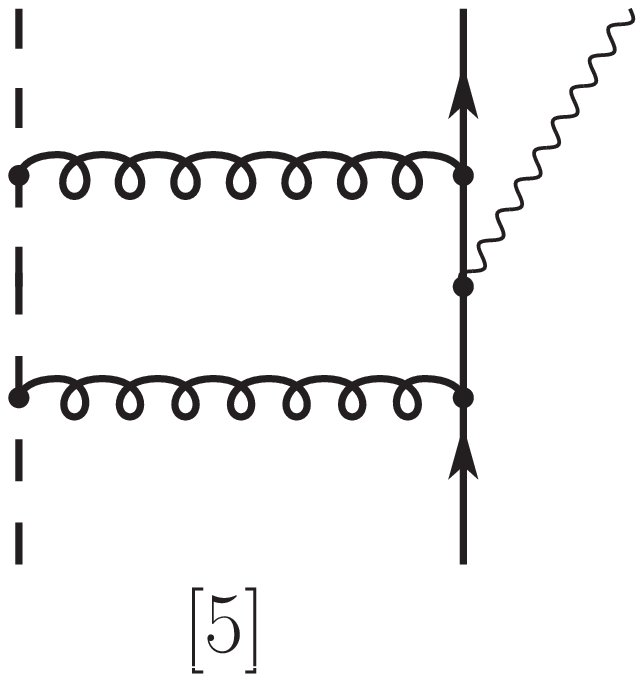,width=30mm}
\end{minipage}
\begin{minipage}[t]{34mm}
\epsfig{file=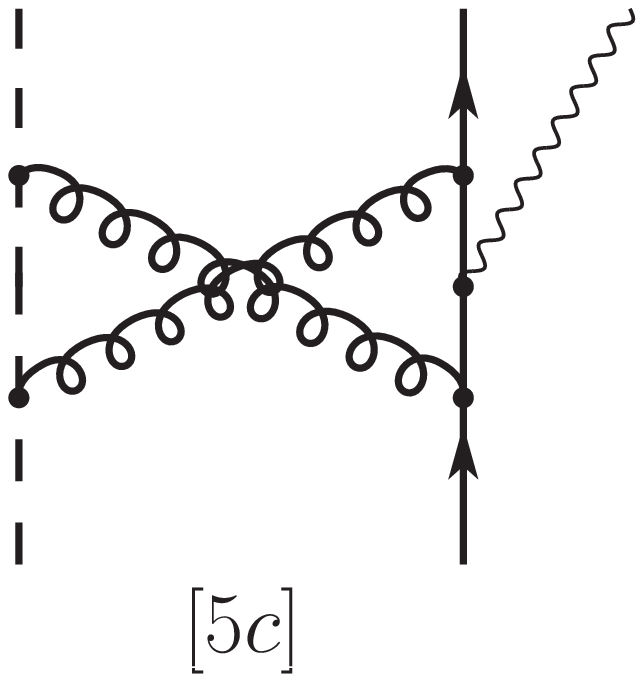,width=30mm}
\end{minipage}
\begin{minipage}[t]{34mm}
\epsfig{file=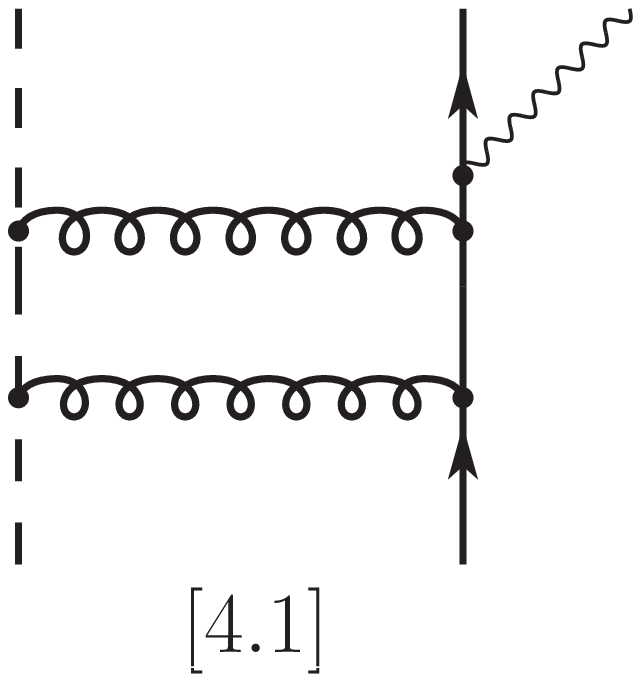,width=30mm}
\end{minipage}
\begin{minipage}[t]{34mm}
\epsfig{file=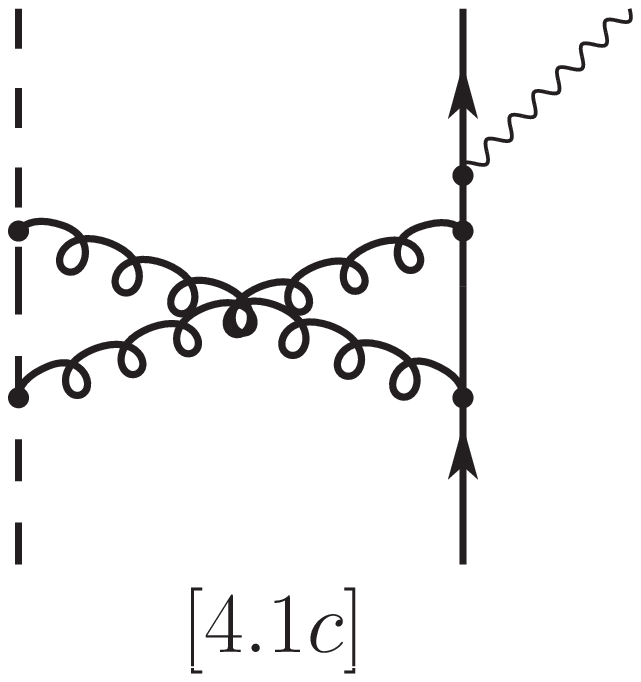,width=30mm}
\end{minipage}
\begin{minipage}[t]{34mm}
\epsfig{file=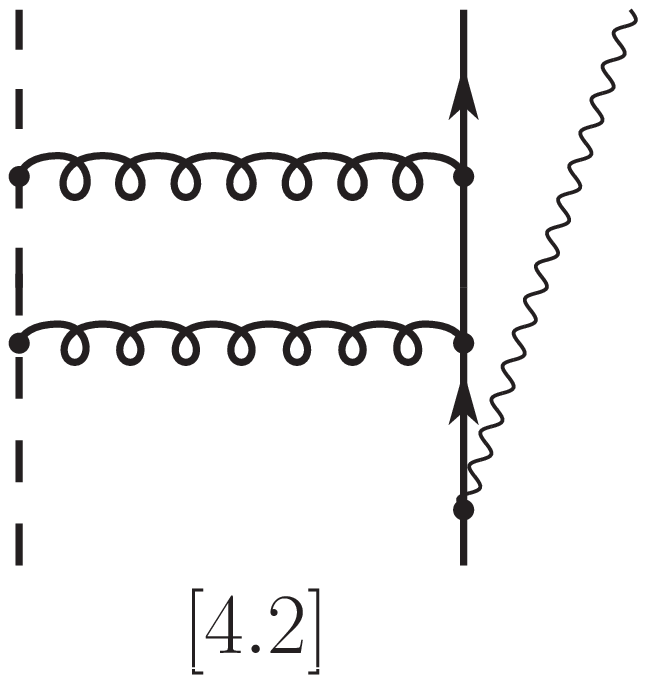,width=30mm}
\end{minipage}
\begin{minipage}[t]{34mm}
\epsfig{file=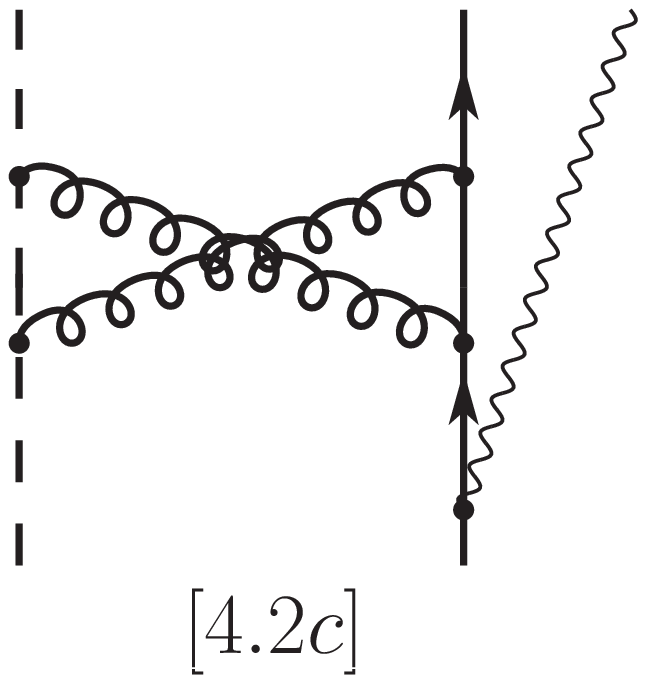,width=30mm}
\end{minipage}
\begin{minipage}[t]{34mm}
\epsfig{file=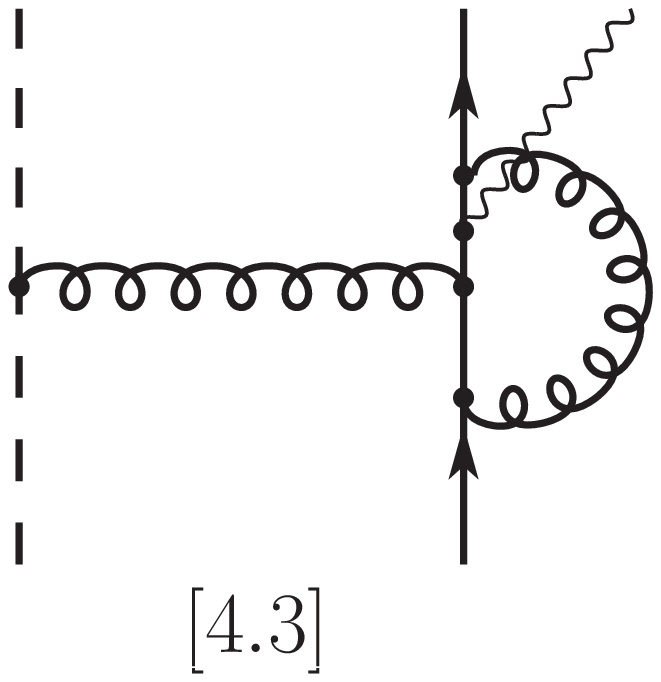,width=30mm}
\end{minipage}
\begin{minipage}[t]{34mm}
\epsfig{file=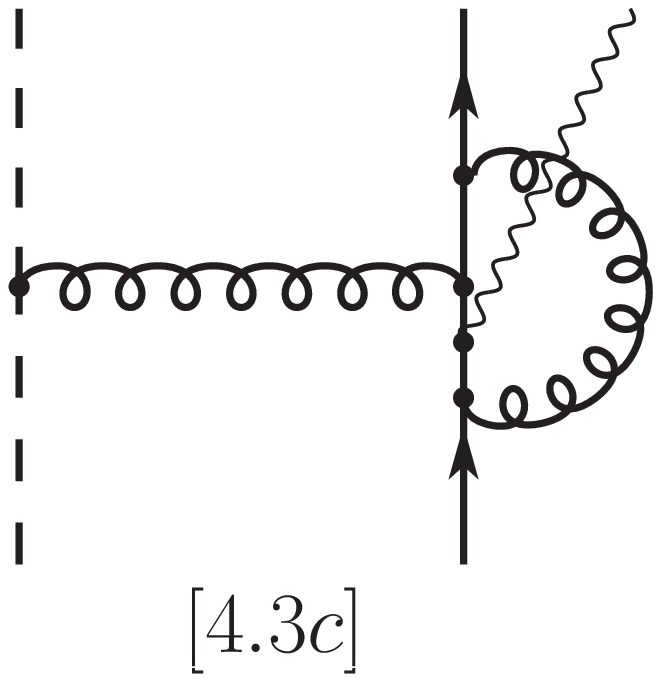,width=30mm}
\end{minipage}
\begin{minipage}[t]{34mm}
\epsfig{file=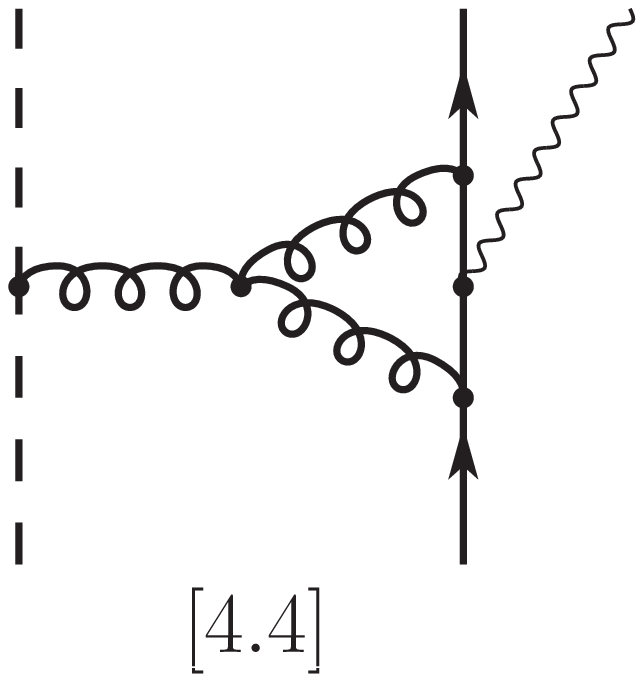,width=30mm}
\end{minipage}
\begin{minipage}[t]{34mm}
\epsfig{file=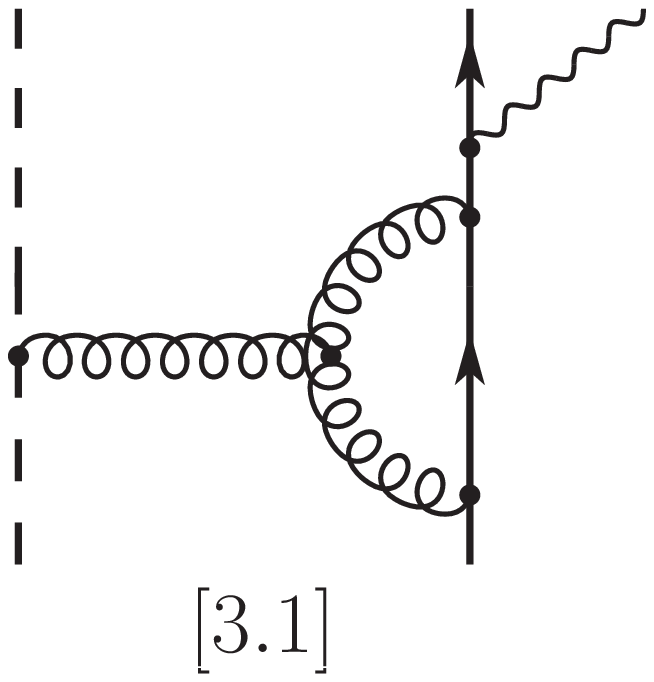,width=30mm}
\end{minipage}
\begin{minipage}[t]{34mm}
\epsfig{file=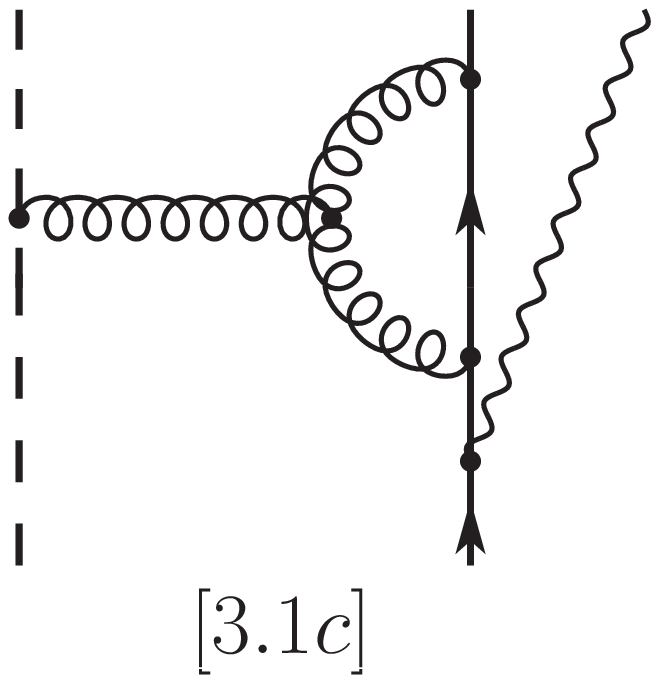,width=30mm}
\end{minipage}
\begin{minipage}[t]{34mm}
\epsfig{file=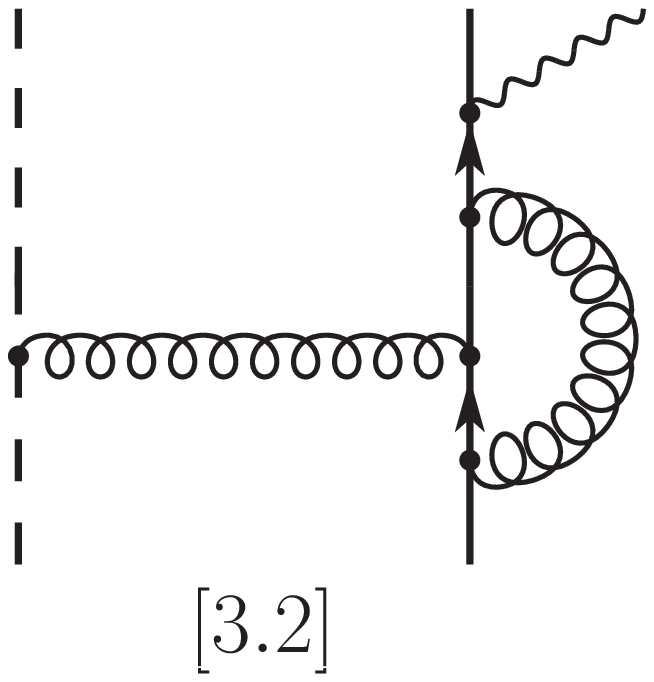,width=30mm}
\end{minipage}
\begin{minipage}[t]{34mm}
\epsfig{file=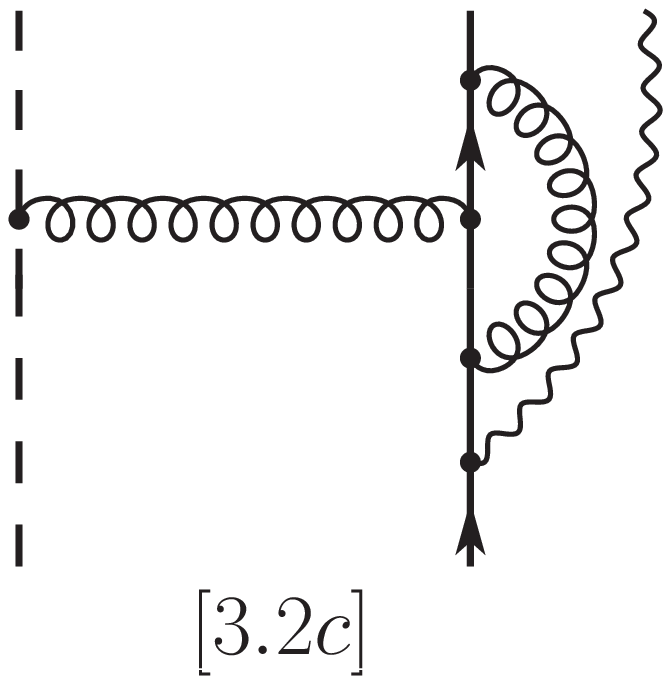,width=30mm}
\end{minipage}
\begin{minipage}[t]{40mm}
\epsfig{file=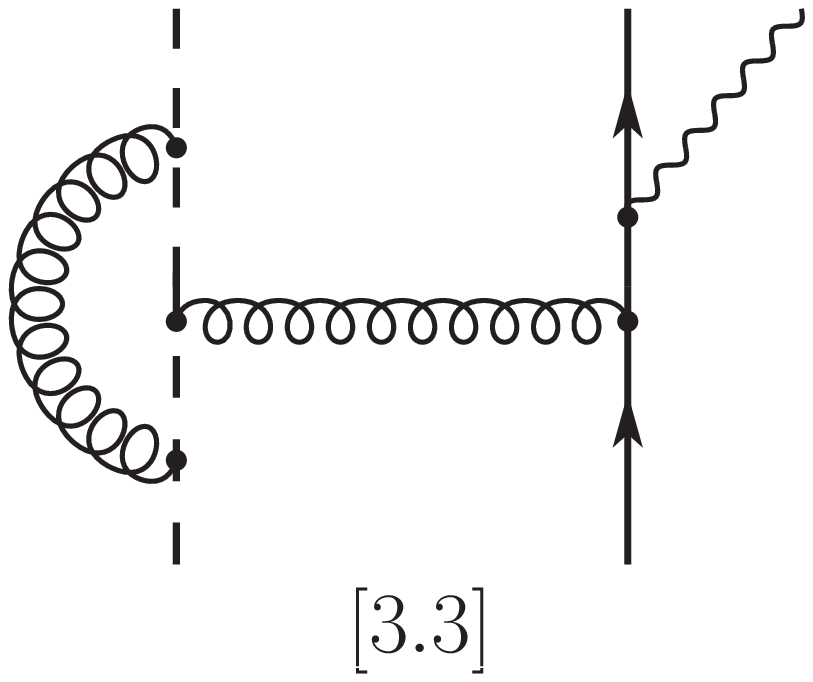,width=39mm}
\end{minipage}
\begin{minipage}[t]{40mm}
\epsfig{file=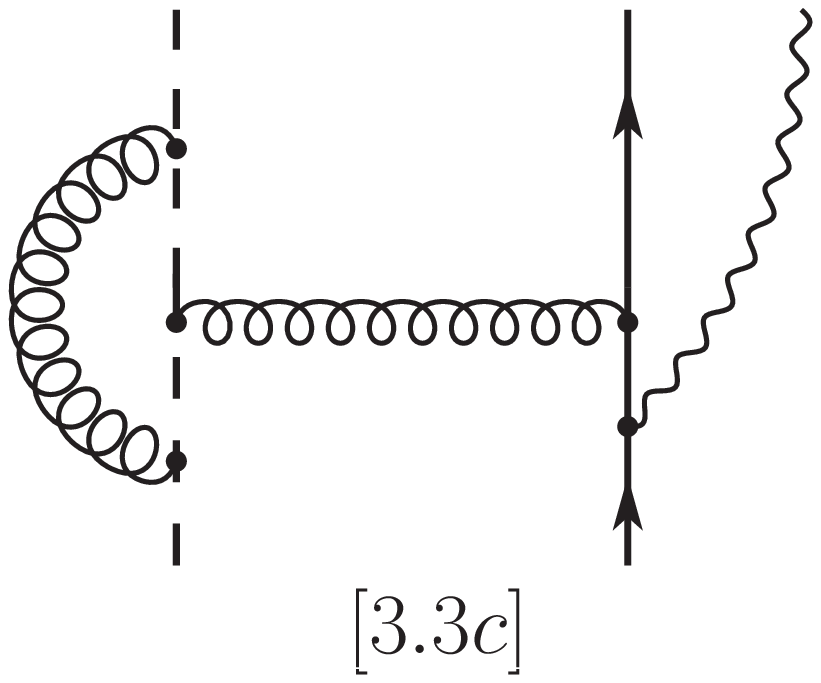,width=39mm}
\end{minipage}
\begin{minipage}[t]{34mm}
\epsfig{file=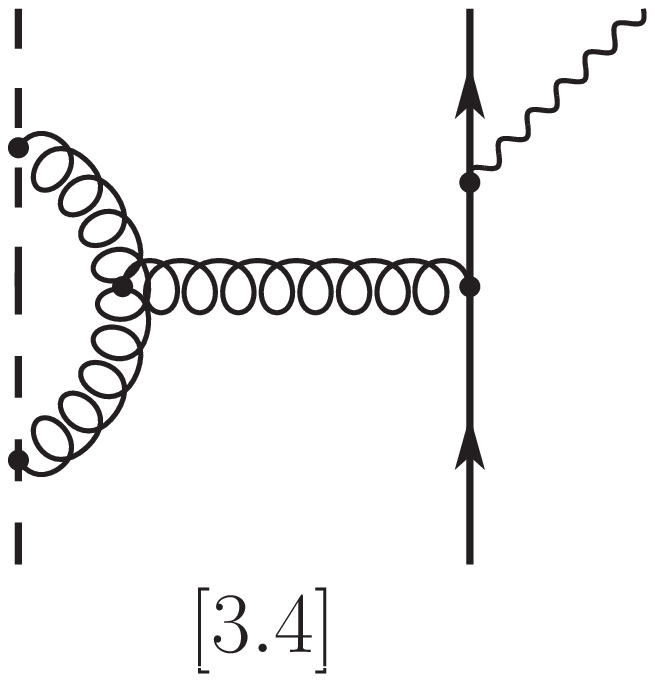,width=30mm}
\end{minipage}
\begin{minipage}[t]{34mm}
\epsfig{file=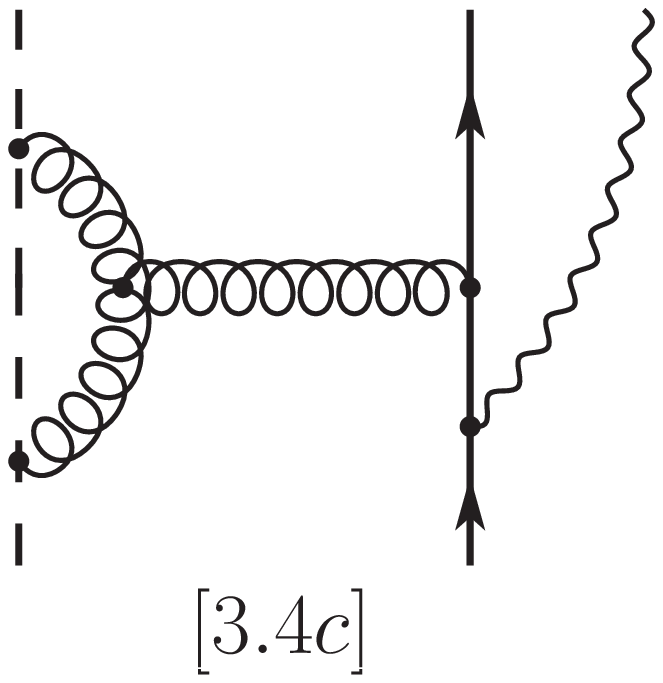,width=30mm}
\end{minipage}
\begin{minipage}[t]{34mm}
\epsfig{file=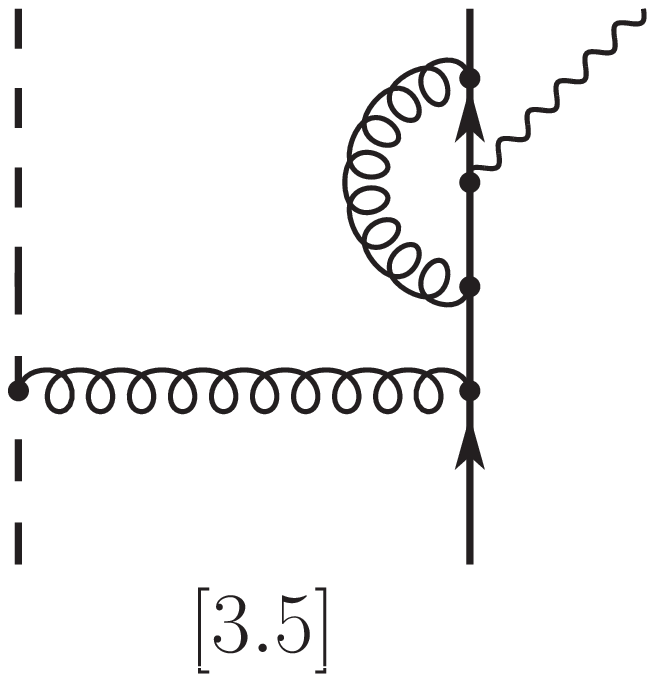,width=30mm}
\end{minipage}
\begin{minipage}[t]{34mm}
\epsfig{file=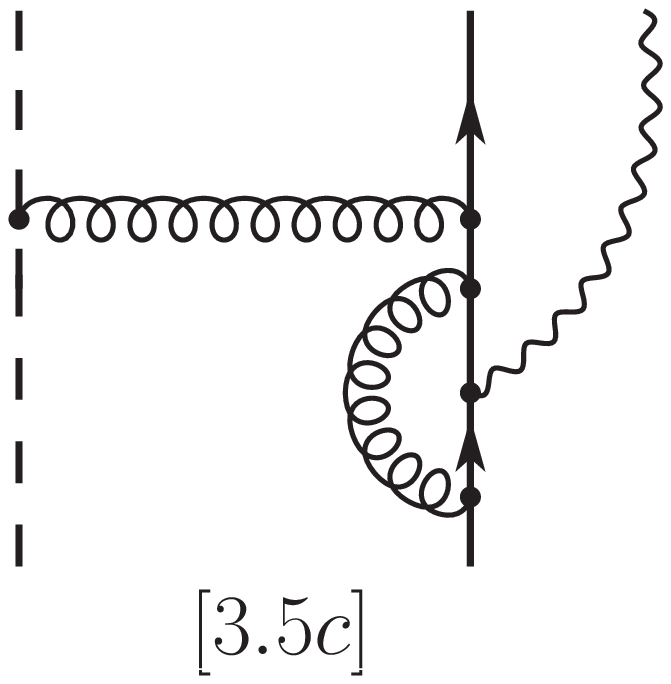,width=30mm}
\end{minipage}
\begin{minipage}[t]{34mm}
\epsfig{file=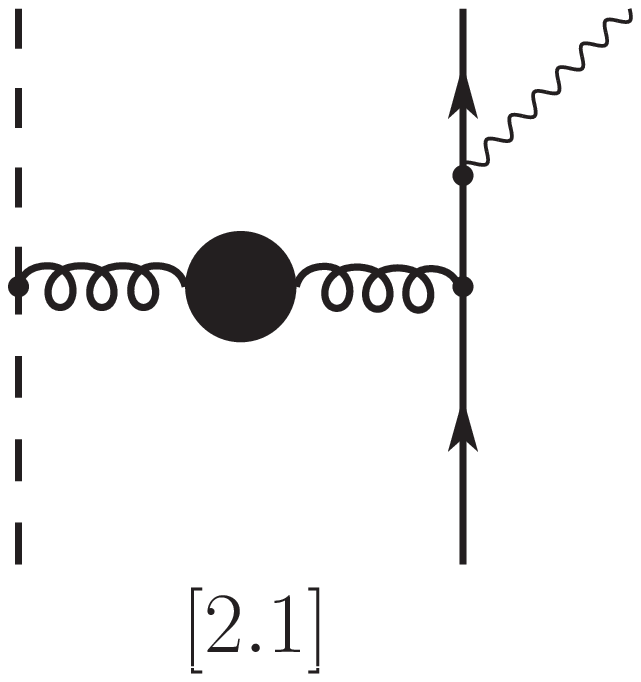,width=30mm}
\end{minipage}
\begin{minipage}[t]{34mm}
\epsfig{file=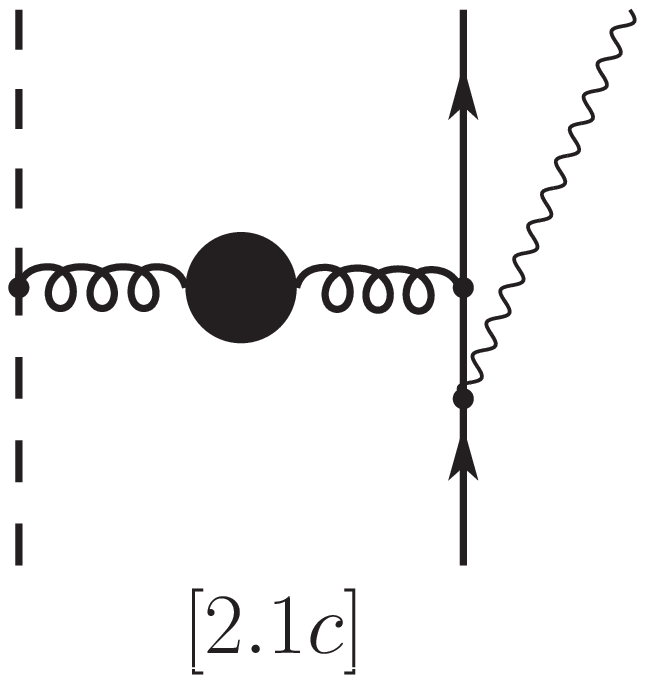,width=30mm}
\end{minipage}
\begin{minipage}[t]{34mm}
\epsfig{file=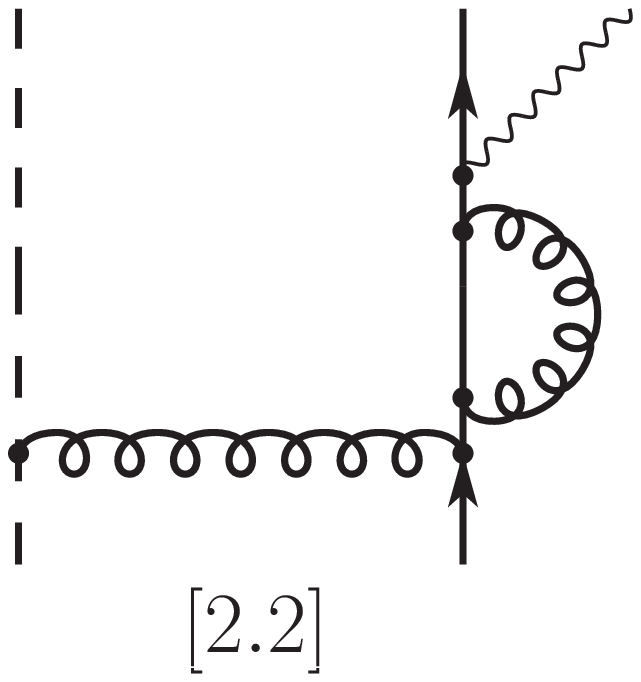,width=30mm}
\end{minipage}
\begin{minipage}[t]{34mm}
\epsfig{file=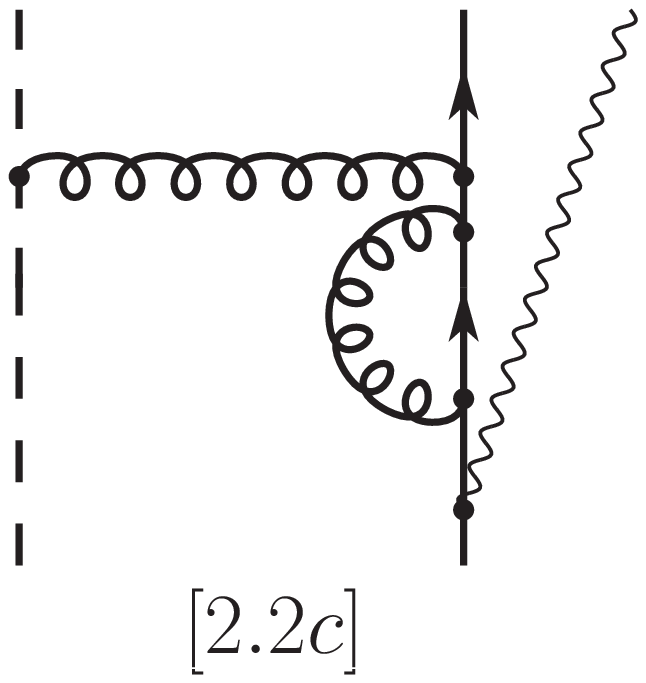,width=30mm}
\end{minipage}
\\[1mm]
\begin{minipage}[t]{16cm}
\epsfig{file=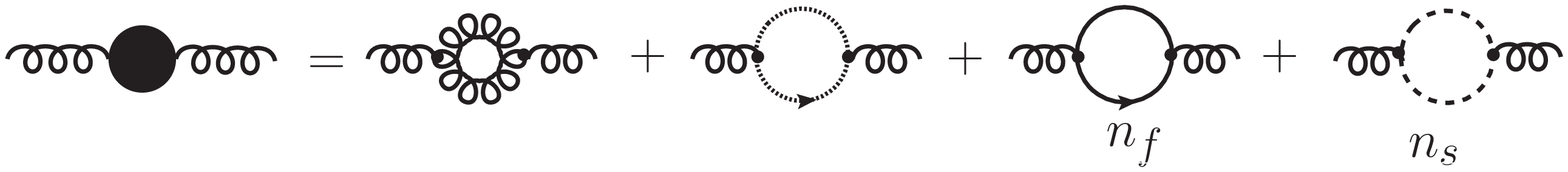,width=150mm}
\end{minipage}
\end{center}
\caption{Nontrivial one-loop diagrams for the process $SQ\rightarrow S'Q'\gamma$:
pentagons(labelled as $[5]$ and crossed diagrams $[5c]$), boxes ($[4.X]$ and crossed
diagrams $[4.Xc]$), triangles ($[3.X]$ and crossed diagrams $[3.Xc]$), and bubbles
($[2.X]$ and crossed diagrams $[2.Xc]$). The last line is the self-energy insertion
with gluons, ghosts, $n_f$ sorts of fermions, and $n_s$ sorts of scalars.}
\label{fig:2}
\end{figure}

We consider all one-loop Feynman diagrams contributing to the process $SQ\rightarrow
S'Q'\gamma$ at first. There are twenty three different nontrivial one-loop Feynman
diagrams: see Fig.~ \ref{fig:2}. There are diagrams labelled according those
belonging to pentagons, boxes, triangles and bubble diagram class. First of all we
perform the reduction of the amplitude to the master integrals by the
``LiteRed''\cite{LiteRed} Mathematica package (author R.N.~Lee: $www.inp.nsk.su/\sim
lee/programs/LiteRed/$). As a result of the reduction one has pentagon, box, and
self-energy nontrivial master integrals. Our master integrals are listed in the
Appendix. The method of our tensor integral calculation is presented in Appendix as
well. The algorithm of the master integral calculation is presented in
Ref.~\cite{BDK}. The next stage is to take Regge limit of the resulting expression
for  master integrals analytically continued in the physical kinematic region.
Finally we compare the result of our calculation with one-loop expression resulting
from the hypothetical multi-Regge form of the amplitude in question. In such a way
we extract the required vertex.

\subsection{Kinematics, color and Lorentz structures of the amplitude}

Momentum of the initial scalar $S$ is $k_1$, of the final scalar $S'$ --- $k_2$, of
the final quark $Q'$ produced in the central rapidity region --- $k_3$, of the final
photon $\gamma$ --- $k_4$, and of the initial quark $Q$ --- $k_5$. In
Fig.~\ref{fig:1} all momenta  are considered to be incoming and lightcone:
$k_1+k_2+k_3+k_4+k_5=0,\quad k_i^2=0$. For the final photon we use the physical
gauge  with a light cone vector along $k_1$:
$\bigl(e(k_4),k_4\bigr)=0,\;\bigl(e(k_4),k_1\bigr)=0$.

We present  Sudakov's decomposition for our momenta with incoming scalar and quark
momenta being along lightcone momenta $n_1$, $n_2$ ($n_1^2=0,\;n_2^2=0,\;
(n_1,n_2)=1$): $k_1=k_1^+n_1$, $k_i=k_i^+n_1+k_i^-n_2+k_{i\bot}$ for $i=2,3,4$, and
$k_5=k_5^-n_2$. Here $n_1$, $n_2$ are light-cone momenta, where
$k_i^{\pm}=(k_i,n_{2,1})$. Here and below the $\bot$ sign is used for components of
momenta transverse to $n_1,\,n_2$ plane. The scalar productions of particle momenta
are expressed through Lorentz invariants:
\begin{equation}
\begin{split}
\label{invariants} &s=2(k_1,k_5),\; t_1=2(k_1,k_2),\;
t_2=2(k_4,k_5),\;s_1=2(k_2,k_3),\;s_2=2(k_3,k_4),\\&
u_1=2(k_1,k_3),\;u_2=2(k_3,k_5),\;u=2(k_1,k_4),\;s'=2(k_2,k_4),\;u'=2(k_2,k_5).
\end{split}
\end{equation}
And we express the other invariants through independent set: $
u_1=t_2-t_1-s_1$,$\;u_2=t_1-t_2-s_2$, $\;u=s_1-t_2-s$,$\;s'=s-s_1-s_2$,
$\;u'=s_2-t_1-s$. It is important that the invariants have the following signs in
the physical region of our process: $s_1>0,\;s_2>0,\;s>0,\;t_1<0,\;t_2<0$ (and
$u_1<0,\;u_2<0,\;u<0,\;s'>0,\;u'<0$).

The multi-Regge kinematics (MRK) means that we have particles well separated in
rapidity space in the final state: \be k_2^+\gg k_3^+\gg k_4^+,\quad k_2^-\ll
k_3^-\ll k_4^-. \ee Since momenta $k_i$ are on the mass shell one has
$k_i^-=-\frac{k_{i\bot}^2}{2k^+_i}\;,\quad i=2,3,4$. We introduce  two dimensionless
large (in the Regge limit) parameters: $y_1=k_2^+/k_3^+ \gg1 $ and $y_2=k_3^+/k_4^+
\gg 1$.  In what follows we will use dimensional regularization $D=4+2\epsilon$
taking the limit $\epsilon \rightarrow 0$ before the Regge limit. MRK applies also
that all  transverse momenta are not increasing as $y_i\rightarrow\infty$. One can
express all of the transverse scalar productions through independent ones
($k_{2\bot}^2,\,k_{3\bot}^2,\,k_{4\bot}^2$):
$2(k_2,k_3)_{\bot}=k_{4\bot}^2-k_{2\bot}^2-k_{3\bot}^2$,
$2(k_3,k_4)_{\bot}=k_{2\bot}^2-k_{3\bot}^2-k_{2\bot}^2$, and
$2(k_2,k_4)_{\bot}=k_{3\bot}^2-k_{2\bot}^2-k_{4\bot}^2$.

Further we consider one-loop amplitude $SQ\rightarrow S'Q'\gamma$ as a power
function of $y_1\,,y_2$ within the accuracy of logarithmic terms (i.e. terms
$\ln^k[y_i]$ technically originating from the $\epsilon$ decomposition of master
integrals). In the Regge limit the leading amplitude behavior is expected to be
$\sim y_1\sqrt{y_2}$. Here $y_1$ comes from Reggeized gluon in the $s_1$-channel and
$\sqrt{y_2}$ ~--- from Reggeized quark in the $s_2$-channel: see Fig.~\ref{fig:1}.
Our basis bispinor structures (\ref{structures-1}) are proportional to $\sqrt{y_2}$,
that is why the order of the multi-Regge limit calculation for the amplitude is as
follows: expansion in $\epsilon\rightarrow0$ with accuracy ${\cal O}(\epsilon)$
followed by retaining the leading asymptotic power expansion in
$y_i\rightarrow\infty$.

There are only two independent color structures: ``tree'' structure $T_{S'S}^at^a$
and ``cross-box'' structure $T^a_{S'c}T^b_{cS}t^bt^a$. Here $t^a$ are $SU(N_c)$
quark generators in the fundamental representation, and $T^a_{S'S}=-i f ^{a\,S'\,S}$
are generators of scalars in the adjoint representation. The ``tree'' structure
(i.e. color octet in $t_1$ channel) turns out to give the leading contribution to
the real part of our amplitude. Next, we use the following notation for the Casimir operator:
\begin{equation}
t^at^a=C_F=\frac{N_c^2-1}{2N_c}\,.
\end{equation}

The amplitude depends not only on the invariants of $s_1,\,s_2,\,t_1,\,t_2,\,s$ but
on the helicity states of the external particles as well.

We assume that external momenta $ k_i $ are embedded in a four-dimensional subspace
of the momentum space with $D=4+2\epsilon$ dimensions, while the photon polarization
is $D$-vector. In that case there are six independent Lorentz helicity structures:
\begin{equation}
\begin{split}
&\bar{u}(k_3)\not{e}u(k_5)\,,\;\bar{u}(k_3)\not{k}_1\not{k}_4\not{e}u(k_5)\,,\;(e,k_2)\bar{u}(k_3)\not{k}_1u(k_5)\,,\;(e,k_2)\bar{u}(k_3)\not{k}_4u(k_5)\,,\\
&(e,k_3)\bar{u}(k_3)\not{k}_1u(k_5)\,,\;(e,k_3)\bar{u}(k_3)\not{k}_4u(k_5)\,.
\label{structures-1}
\end{split}
\end{equation}
In the multi-Regge kinematics we can choose the following independent structures
with only transverse ($D-2$) components involved:
\begin{equation}
\begin{split}
&\bar{u}(k_3)\not{e}_{\bot}u(k_5)\,,\;\bar{u}(k_3)\not{k}_{2\bot}\not{k}_{4\bot}\not{e}_{\bot}u(k_5)\,,\;(e,k_2)_{\bot}\bar{u}(k_3)\not{k}_{2\bot}u(k_5)\,,\;(e,k_2)_{\bot}\bar{u}(k_3)\not{k}_{4\bot}u(k_5)\,,\\
&(e,k_4)_{\bot}\bar{u}(k_3)\not{k}_{2\bot}u(k_5)\,,\;(e,k_4)_{\bot}\bar{u}(k_3)\not{k}_{4\bot}u(k_5)\,.
\label{structures-2}
\end{split}
\end{equation}
In the limit $D\rightarrow4$ the first two structures become dependent since one can
express $\not{e}_{\bot}$ in terms of $\not{k}_{2\bot}, \, \not{k}_{4\bot}$:
\begin{equation}
\begin{split}
e_{\bot}^{\mu}=k_{2\bot}^{\mu}\frac{k_{4\bot}^2(e,k_2)_{\bot}-
(k_2,k_4)_{\bot}(e,k_4)_{\bot}}{k_{2\bot}^2k_{4\bot}^2-(k_2,k_4)^2_{\bot}}+
k_{4\bot}^{\mu}\frac{-(k_2,k_4)_{\bot}(e,k_2)_{\bot}+k_{2\bot}^2(e,k_4)_{\bot}}{k_{2\bot}^2k_{4\bot}^2-
(k_2,k_4)^2_{\bot}}.
\end{split}
\end{equation}
This means that  the part of metric tensor $g^{\mu\nu}_{D-4}\sim {\cal O}(\epsilon)$
vanishes in the dimensional limit $D\rightarrow4$.

\section{Regge amplitude structure}

According to the hypothesis of the quark and gluon Reggeization in NLA the real part of the amplitude $A+B\rightarrow A'+J_1+\dots+J_n+B'$ in the MRK has the form
\begin{equation}\label{A 2-2+n}
\Re {\cal A}_{2\rightarrow n+2}=\bar{\Gamma}^{{R}_1}_{A^{\prime}A} \left({\cal P}\prod_{i=1}^n e^{\omega_{R_i}(q_i)(z_{i-1}-z_i)}\widehat{D}_{R_i}\gamma^{J_i}_{{R}_i {R}_{i+1}}\right) e^{\omega_{R_{n+1}}(q_{n+1})(z_{n}-z_{n+1})} \widehat{D}_{R_{n+1}} \Gamma^{{R}_{n+1}}_{ B^{\prime} B}\,,
\end{equation}
where ${\cal P}\prod$ is the product ordered along the fermion line. We use the notation
\begin{equation}
\widehat{D}_{R_i}=
\begin{cases}
&\dfrac{1}{q_{i\bot}^2}\,,\;R_i={\cal G}_i, \\
&-\dfrac{\not{q}_{i\bot}}{q_{i\bot}^2}\,,\;R_i={\cal Q}_i,
\end{cases}
\end{equation}
for the Reggeon $R_i$ (gluon or quark) propagator. Then $z_i=\frac{1}{2}\ln\frac{k_i^+}{k_i^-}$ are rapidities of the final jets $J_i$. In next-to-leading logarithmic approximation (NLA) jets $J_i$ are either one parton or the partons couple with close rapidities. 
Lastly, $\omega_{R}(q)$ in (\ref{A 2-2+n}) is the Regge trajectory of the Reggeon $R$ (gluon or quark) with momentum $q$.

There are several effective vertices ($\bar{\Gamma}_{A'A}^{R_1},\,\Gamma_{B'B}^{R_{n+1}}$) for the particle--jet transition in the ``fragmentation'' kinematic region. As for particle--particle transition (pure multi-Regge kinematics) one has the following vertices in QCD:
\[
\Gamma_{G'G}^{{\cal G}}\,,\;\Gamma_{Q'Q}^{{\cal G}}\,,\;\Gamma_{G'Q}^{{\cal Q}}.
\]
For SYM there is an extra vertex $\Gamma_{S'S}^{\cal G}$. All these vertices are calculated within the NLO~\cite{FF:2001} (for SYM case see \cite{FKR:2013}). The vertex the quark-photon transition $\Gamma_{\gamma' Q}^{Q_2}$ may be found in \cite{FF:2001} as well. For quasi-multi-Regge kinematics (QMRK) case of particle--couple transition  one has
\[
\Gamma_{\{G_1G_2\}G}^{{\cal G}}\,,\;\Gamma_{\{Q_1Q_2\}G}^{{\cal G}}\,,\;\Gamma_{\{G_1Q_2\}Q}^{{\cal G}}\,,\;\Gamma_{\{G_1G_2\}Q}^{{\cal Q}}\,,\;\Gamma_{\{G_1Q_2\}G}^{{\cal Q}}\,,\;\Gamma_{\{Q_1Q_2\}Q}^{{\cal Q}}
\]
in QCD. All these vertices are known: see ~\cite{LV:2001,F:2003}. 
For SYM there are some additional vertices ($\Gamma_{\{S_1S_2\}G}^{\cal G}$, $\;\Gamma_{\{G S'\}S}^{\cal G}$, $\Gamma_{\{Q_1Q_2\}S}^{\cal G}$, $\;\Gamma_{\{Q'S\}Q}^{\cal G} $): the corresponding calculation one can find in Ref.~\cite{FKR:2013}.

There are several  effective vertices ($\gamma_{R_iR_{i+1}}^{J_i}$) for the jet production in Reggeon-Reggeon collision in the central region of rapidity. For one-particle production we have QCD vertices
\[
\gamma_{{\cal G}_1{\cal G}_2}^{G}\,,\;\gamma_{{\cal Q}_1{\cal Q}_2}^{G}\,,\;\gamma_{{\cal Q}_1{\cal G}_2}^{Q}\,.
\]
Vertices $\gamma_{{\cal G}_1{\cal G}_2}^{G}$, $\gamma_{{\cal Q}_1{\cal Q}_2}^{G}$ were calculated in NLO~\cite{FL:1993,FFP:2001,DS:1999,BG:2007}. Effective vertex $\gamma_{{\cal Q}_1{\cal G}_2}^{Q}$ was calculated in the leading order only. Our purpose is to find one-loop corrections to it.   Vertices for couple production (QMRK) in the central region of rapidity
\[
\gamma_{{\cal G}_1{\cal G}_2}^{\{G_1G_2\}}\,,\;\gamma_{{\cal G}_1{\cal G}_2}^{\{Q_1Q_2\}}\,,\;\gamma_{{\cal Q}_1{\cal G}_2}^{\{Q_1G_2\}}\,,\;\gamma_{{\cal Q}_1{\cal Q}_2}^{\{G_1G_2\}}\,,\;\gamma_{{\cal Q}_1{\cal Q}_2}^{\{Q_1Q_2\}}\,
\]
are calculated in QCD with required NLO accuracy~\cite{LV:2001,F:2003} as well. In SYM one has an extra vertex $\gamma_{R_1R_2}^{\{S_1S_2\}}$ that was calculated some years ago \cite{FG:2008}.

Note that QMRK is subleading kinematic i.e. all necessary vertices are on the tree level. For QMRK amplitudes with gluon and quark exchanges  the Reggeization hypothesis  was proved in~\cite{BG:2006}.

In the following we will use the standard momentum notations for the Regge amplitude $SQ\rightarrow S'Q'\gamma$ (see Fig.~\ref{fig:1}):
\begin{equation}
q_{1\bot}\equiv (k_2+k_1)_{\bot} =k_{2\bot},\;q_{2\bot}\equiv (-k_4-k_5)_{\perp} = -k_{4\bot},\;k_{\bot}\equiv -k_{3\bot},
\end{equation}
with $k$ being momentum of the quark produced.
According to the hypothesis of the quark and gluon Reggeization in NLA (\ref{A 2-2+n}), the real
part of the amplitude $SQ\rightarrow S'Q'\gamma $ in the multi-Regge kinematics reads as
\begin{equation}\label{eq:Rsq}
\Re {\cal A}_8=\Gamma_{S'S}^{R_1}\biggl(\frac{s_1}{\sqrt{
q_{1\bot}^2k_{\bot}^2}}\biggr)^{\omega_g(q_1)}\frac{1}{q_{1\bot}^2}\gamma_{R_1Q_2}^{Q}\biggl
(\frac{s_2}{\sqrt{k_{\bot}^2q_{2\bot}^2}}\biggr)^{\omega_q(q_2)}
\biggl[-\frac{\not{q}_{2\bot}}{q_{2\bot}^2}\biggr]\Gamma_{\gamma'Q}^{Q_2}.
\end{equation}

\subsection{Regge trajectories and effective vertices}
Now we present an expression for the one-loop trajectories of a quark and gluon:
\begin{equation}
\omega_q(q)=-2C_Fg^2(-ia_{\Gamma})\frac{(-q_{\bot}^2)^{\epsilon}}{\epsilon}\,,
\;\omega_g(q)=-2N_cg^2(-ia_{\Gamma})\frac{(-q_{\bot}^2)^{\epsilon}}{\epsilon}.
\end{equation}
The constant $a_{\Gamma}$ emerges from the integrals as a common factor
(\ref{eq:aG}). The combination of $N_cg^2(-ia_{\Gamma})$ will arise often and  it
relates with $\bar{g}^2$ notation as follows:
\begin{equation}
N_cg^2(-ia_{\Gamma})\equiv N_cg^2\frac{\Gamma(1-\epsilon)}{(4\pi)^{2+\epsilon}}
\frac{\Gamma^2(1+\epsilon)}{\Gamma(1+2\epsilon)}=\bar{g}^2\Bigl(1-\frac{\pi^2}{6}\epsilon^2+{\cal
O}(\epsilon^3)\Bigr).
\end{equation}

Now we present the scalar to scalar Regge vertex in a following way
~\cite{FKR:2013}:
\begin{equation}
\Gamma_{S'S}^{R_1}=2k_1^+gT_{S'S}^{R_1}(1+\delta_S)\;, \quad
\delta_S=\delta_S^{c}+\delta_S^{s.e.}+\delta_{S}^{v}+\delta_S^{A}. \label{GammaSS}
\end{equation}
Expressions for the corrections ($\delta$'s) to vertex of the scalar scattering can
be found in ~\cite{FKR:2013} (there are corrections in the framework of
supersymmetric Yang--Mills theory, but the QCD result can be obtained easily):
\begin{equation}
\delta_{S}^{A}+\delta_{S}^{c}=(-ia_{\Gamma})g^2N_c
(-q_{1\bot}^2)^{\epsilon}\biggl(-\frac{5}{4\epsilon^2}+\frac{1}{2\epsilon}-1+\frac{\pi^2}{2}\biggr),
\end{equation}
\begin{equation}
\delta_S^{s.e.}=(-ia_{\Gamma})N_cg^2\frac{(-q_{1\bot}^2)^{\epsilon}}
{\epsilon}\biggl(-\Bigl[\frac{5}{6}-\frac{31}{18}\epsilon\Bigr]+
n_s\Bigl[\frac{1}{12}-\frac{2}{9}\epsilon\Bigr]+\frac{n_f}{N_c}\Bigl[
\frac{1}{3}-\frac{5}{9}\epsilon\Bigr]\biggr),
\end{equation}
\begin{equation}
\delta_S^{v}=(-ia_{\Gamma})N_cg^2\frac{(-q_{1\bot}^2)^{\epsilon}}
{\epsilon^2}\biggl(\Bigl[\frac{5}{4}-\frac{3}{2}\epsilon+3\epsilon^2\Bigr]+\Bigl[-2+4\epsilon-8\epsilon^2\Bigr]\biggr).
\end{equation}
Here the superscript $c$ denotes the universal  contribution from the central
rapidity region, the index $s.e.$ denotes the contribution of the mass (self-energy)
operator, the index $v$ denotes the contribution of the vertex corrections, and the
index $A$ represents the contribution coming from the rapidity close to the initial
particle. All corrections are presented in $\epsilon$ decomposition with required
accuracy.

Corrections to vertex of photon production is more complicated since the
structure contains helicity violating terms: 
\begin{equation}
\Gamma_{\gamma' Q}^{Q_2}=-e\bigl(\not{e}_{\bot}+\not{e}_{\bot}\delta_{1\gamma}+
\frac{(eq_2)_{\bot}}{q_{2\bot}^2}\not{q}_{2\bot}\delta_{2\gamma}\bigr)u(k_5),
\label{GammagammaQ}
\end{equation}
\begin{equation}
\delta_{1\gamma}=\delta_{1\gamma}^{s.e.}+\delta_{1\gamma}^{v}+\delta_{1\gamma}^{A}+\delta_{1\gamma}^{c}\,,\quad
\delta_{2\gamma} =\delta_{2\gamma}^{v}+\delta_{2\gamma}^A.
\end{equation}
Expression for these corrections can be found  in~\cite{FF:2001}:
\begin{equation}
\delta_{1\gamma}^{A}+\delta_{1\gamma}^{c}=g^2(-ia_{\Gamma})
(-q_{2\bot}^2)^{\epsilon}(-C_F)\biggl[\frac{1}{\epsilon^2}-\frac{\pi^2}{2}\biggr],
\end{equation}
\begin{equation}
\delta_{1\gamma}^v=C_Fg^2(-ia_{\Gamma})(-q_{2\bot}^2)^{\epsilon}\frac{1-4\epsilon}{\epsilon},
\end{equation}
\begin{equation}
\delta_{1\gamma}^{s.e.}=C_Fg^2(-ia_{\Gamma})(-q_{2\bot}^2)^{\epsilon}\frac{1-\epsilon}{2\epsilon},
\end{equation}
\begin{equation}
\delta_{2\gamma}^{v}=C_Fg^2(-ia_{\Gamma})(-q_{2\bot}^2)^{\epsilon}\frac{(-2)(2-5\epsilon)}{\epsilon},
\end{equation}
\begin{equation}
\delta_{2\gamma}^{A}=C_Fg^2(-ia_{\Gamma})(-q_{2\bot}^2)^{\epsilon}\frac{4(1-2\epsilon)}{\epsilon}.
\end{equation}

We parametrize the unknown vertex of quark production in quark-Reggeon collision as 
\begin{equation}
\gamma_{R_1Q_2}^{Q}=-g\frac{1}{k_3^+}\bar{u}(k_3)t^{R_1}\bigl(\not{q}_{1\bot}+
\not{q}_{1\bot}\delta_{1Q}+\not{q}_{2\bot}\delta_{2Q}\bigr).
\end{equation}
The term $\delta_{1Q}$ is a correction to leading order structure. The correction
$\delta_{2Q}$ stands before the structure, that is absent in the leading order (the
mass operator corrections contribute only in the $\delta_{1Q}$ coefficient):
\begin{equation}
\delta_{1Q}=\delta_{1Q}^{s.e.1}+\delta_{1Q}^{s.e.2} +\delta_{1Q}^{v,c}\,, \quad
\delta_{2Q}=\delta_{2Q}^{v,c}.
\end{equation}

\subsection{Lorentz and color structures of the Regge amplitude}
Now we consider the real part of the amplitude in question.

In the first place we will be interested in octet color, or ``tree'', structure
coefficient of Regge amplitude obtained in our calculation after the Regge limit
procedure.

Let us introduce the notation for the basic Lorentz structures of our Regge
amplitude. There is only one Born structure
\begin{equation}
A^{Born}=-2y_1g^2eT_{S'S}^{R_1}\frac{\bar{u}(k_3)t^{R_1}\not{q}_{1\bot}\not{q}_{2\bot}
\not{e}_{\bot}u(k_5)}{q_{1\bot}^2q_{2\bot}^2}. \label{structureABorn}
\end{equation}
The next structure $A_8^{e}$ arises from the correction to the Regge vertex for
quark production in the central region and violates the helicity
\begin{equation}
A^{e}_8=-2y_1g^2eT_{S'S}^{R_1}\frac{\bar{u}(k_3)t^{R_1}\not{e}_{\bot}u(k_5)}{q_{1\bot}^2}.
\end{equation}
Structure $A^{q_1}_8$ arises from correction to the Regge  vertex for quark-photon
transition and violates the helicity as well
\begin{equation}
A^{q_1}_8=-2y_1g^2eT_{S'S}^{R_1}\frac{\bar{u}(k_3)t^{R_1}\not{q}_{1\bot}u(k_5)(e,q_2)_{\bot}}{q_{1\bot}^2q_{2\bot}^2}.
\end{equation}

The final structure that appears after the Regge limit in our calculations is as
follows
\begin{equation}
A^{q_2}_8=-2y_1g^2eT_{S'S}^{R_1}\frac{\bar{u}(k_3)t^{R_1}\not{q}_{2\bot}u(k_5)(e,q_2)_{\bot}}{q_{1\bot}^2q_{2\bot}^2}.
 \label{structureAq2}
\end{equation}

Decomposition of the multi-Regge form of the amplitude $SQ\rightarrow S'Q'\gamma$ (\ref{eq:Rsq}) in the coupling constant up to the next-to-leading order gives us:
\begin{equation}
\begin{split}
&\Re {\cal A}_8=A^{Born}\Bigl(1+\omega_g(q_1)\ln y_1+\omega_q(q_2)\ln
y_2+\frac{\omega_g(q_1)}{2}
\ln\Big[\frac{k_{\bot}^2}{q_{1\bot}^2}\Big]+\frac{\omega_q(q_2)}{2}\ln\Big[\frac{q_{2\bot}^2}{k_{\bot}^2}\Big]+\\
&+
\delta_S+\delta_{1Q}+\delta_{1\gamma}\Bigr)+A^{e}_8\delta_{2Q}+A^{q_1}_8\delta_{2\gamma}+{\cal
O}(eg^6).
\label{multiReggeform}
\end{split}
\end{equation}
If we calculate the one-loop corrections for $SQ\rightarrow S'Q'\gamma$ amplitude and substract the known corrections for the effective vertices $\Gamma_{S'S}^{\cal G}$, $\Gamma_{\gamma Q}^{\cal Q}$ and the terms with Regge trajectories contribution, then we obtain the corrections for effective vertex $\gamma_{{\cal G}{\cal Q}}^Q$.
\section{Result of the amplitude $SQ\rightarrow S'Q'\gamma$ calculation}
Let us present the result of the calculation procedure described at the beginning of
Section \ref{section-Amplitude}. Here we give the calculation result in the Regge
limit and group diagrams into the expressions with specific elements for the Regge
amplitude. We use notations ( \ref{structureABorn})--(\ref{structureAq2}) for
structures from the previous Section and the notation (\ref{eq:aG}) for the common
factor $a_{\Gamma}$. 

Now we present characteristic diagram contributions reproducing different components of the Regge amplitude: photon and scalar vertex corrections, Regge trajectories, and the corrections to the unknown vertex.

The sum of diagrams giving the correction to the photon vertex reads as
\begin{equation}
\Re\bigl(A_{3.5}+A_{3.5c}\bigr)= g^2(-ia_{\Gamma})\,C_F\biggl\{-A_8^e\,\ln
y_2-(-q_{2\bot}^2)^{\epsilon}\frac{1}{\epsilon}
\Bigl[2(2-5\epsilon)\bigl(A_8^{q_1}+A_8^{q_2}\bigr)-(1-4\epsilon)A^{Born}\Bigr]\biggr\}.
\end{equation}
It is easy to see that  these diagrams (3.5 group) contain the large logarithm $\ln
y_2$. Diagrams describing the mass operator of the quark in the $t_2$-channel
contain  $ \ln y_2 $ as well:
\begin{equation}
\Re\bigl(A_{2.2}+A_{2.2c}\bigr)=g^2(-ia_{\Gamma})\,
C_F\biggl(A^{Born}\frac{1-\epsilon}{\epsilon}(-q_{2\bot}^2)^{\epsilon}-A_e^8\ln
y_2\biggr).
\end{equation}

Diagrams of the vacuum polarization of the  gluon in the $t_1$-channel  result in
the expression (in $\epsilon \rightarrow 0$ decomposition)
\begin{equation}
A_{2.1X}=N_cg^2(-ia_{\Gamma}) A^{Born}\frac{(-q_{1\bot}^2)^{\epsilon}}
{\epsilon}\biggl(-\Bigl(\frac{5}{3}- \frac{31}{9}\epsilon\Bigr)+\frac{n_s}{2}
\Bigl(\frac{1}{3}-\frac{8}{9}\epsilon\Bigr)+
\frac{n_f}{N_c}\Bigl(\frac{2}{3}-\frac{10}{9}\epsilon\Bigr)\biggr).
\end{equation}

The following group of diagrams gives the correction to the scalar vertex:
\begin{equation}
\begin{split}
A_{3.3}+A_{3.3c}+A_{3.4}+A_{3.4c}&=N_cg^2(-ia_{\Gamma})A^{Born}\frac{(-q_{1\bot}^2)^{\epsilon}}{\epsilon^2}\times\\
&\times\biggl(-(2-4\epsilon+8\epsilon^2)+
\frac{1}{4}(5-6\epsilon+12\epsilon^2)\biggr).
\end{split}
\end{equation}

Diagrams describing Reggeization of quarks and gluons (i.e. yielding the Regge
trajectories) and delivering  the correction to the vertex of the quark production
in the central region give the following real part for the octet (tree) color
structure:
\begin{equation}
\begin{split}
&\Re\Bigl(A_{5}+A_{4.1}+A_{4.2}+A_{4.3}+A_{4.3c}+A_{4.4}+A_{3.1}+A_{3.1c}+A_{3.2}+A_{3.2c}\Bigr)\bigg|_{8}=-ig^2a_{\Gamma}\times\\
&\times\Biggl\{
A_{8}^e\Biggl[(N_c-C_F)\biggl(1+\frac{k_{\bot}^2}{q_{1\bot}^2-q_{2\bot}^2}-\frac{q_{1\bot}^2k_{\bot}^2}{\bigl(q_{1\bot}^2-q_{2\bot}^2\bigr)^2}\ln\biggl[\frac{q_{1\bot}^2}{q_{2\bot}^2}\biggr]\biggr)+\\
&+C_F\biggl(\frac{3q_{1\bot}^2}{q_{1\bot}^2-q_{2\bot}^2}\ln\biggl[\frac{q_{1\bot}^2}{q_{2\bot}^2}\biggr]+2\ln y_2\biggr)\Biggr]+A^{Born}\Biggl[C_F\Biggl(-\frac{2-\epsilon}{\epsilon^2}(-q_{2\bot}^2)^{\epsilon}-\frac{2}{\epsilon}(-q_{2\bot}^2)^{\epsilon}\ln y_2+\\
&+\frac{2\pi^2}{3}-4+\frac{3q_{1\bot}^2}{q_{1\bot}^2-q_{2\bot}^2}\ln\biggl[\frac{q_{1\bot}^2}{q_{2\bot}^2}\biggr]+2\Li_2\biggl[1-\frac{q_{1\bot}^2}{q_{2\bot}^2}\biggr]\Biggr)-N_c\Biggl(\frac{1}{\epsilon^2}(-k_{\bot}^2)^{\epsilon}+\frac{2}{\epsilon}(-q_{1\bot}^2)^{\epsilon}\ln y_1+\\
&+\frac{1+2\epsilon}{4\epsilon^2}(-q_{1\bot}^2)^{\epsilon}-\frac{2\pi^2}{3}-2+\ln\biggl[\frac{q_{1\bot}^2}{q_{2\bot}^2}\biggr]\ln\biggl[\frac{k_{\bot}^2}{q_{2\bot}^2}\biggr]+2\Li_2\biggl[1-\frac{q_{1\bot}^2}{q_{2\bot}^2}\biggr]\Biggr)\Biggr]+\\
&+A_8^{q_1}\biggl[4C_F\frac{1-2\epsilon}{\epsilon}(-q_{2\bot}^2)^{\epsilon}\biggr]+A_8^{q_2}\biggl[2C_F\frac{2-5\epsilon}{\epsilon}(-q_{2\bot}^2)^{\epsilon}\biggr]
\Biggr\}.
\end{split}
\end{equation}

\subsubsection*{Corrections to Regge vertices}

Contribution to the mass operator in the $t_1$-channel comes from diagrams $A_{2.1}$ и $A_{2.1c}$
\begin{equation}
A_{2.1}+A_{2.1c}=2\delta_S^{s.e.}A^{Born}=2\delta_{1Q}^{s.e.1}A^{Born}\,.
\label{self-energy contrib-1}
\end{equation}
Nonlogarithmic part of diagrams $A_{2.2}$ and $A_{2.2c}$  (the large  logarithm $\ln
y_2$ presents in the diagram $A_{2.2}$) contributes to the mass operator in the
$t_2$-channel resulting in
\begin{equation}
\Re(A_{2.2}+A_{2.2c})\Big|_{(\ln
y_2)^0}=2\delta_{1Q}^{s.e.2}A^{Born}=2\delta_{1\gamma}^{s.e.}A^{Born}\,.
\label{self-energy contrib-2}
\end{equation}
The sum of the diagrams $3.3X$ and $3.4X$ contributes to the correction
$\delta_S^v$:
\begin{equation}
A_{3.3}+A_{3.3c}+A_{3.4}+A_{3.4c}=A^{Born}\delta_S^{v}.
\end{equation}
The sum of the box-type diagrams in the $t_1$-channel with large logarithms has only
the tree color structure (octet in $t_1$-channel) and yields the gluon trajectory
\begin{equation}
(A_{4.1}+A_{4.2}+A_{4.1c}+A_{4.2c})\Big|_{\ln y_1}=A^{Born}\omega_g(q_1)\ln y_1.
\end{equation}
In the cross-box color structure the large logarithm $\ln y_1 $ cancels completely
(according to the gluon Reggeization).

For $t_2$-channel expression before the large logarithm $\ln y_2$ is reduced to the
quark trajectory by more complex way than in $t_1$-channel:
\begin{equation}
(A_{4.3}+A_{4.3c}+A_{4.4}+A_{3.1}+A_{3.2}+A_{3.5}+ A_{2.2})\Big|_{\ln
y_2}=A^{Born}\omega_q(q_2)\ln y_2.
\end{equation}

Squares of large logarithms $\ln y_2$ cancel in the following sums:
\begin{equation}
(A_{4.3}+A_{4.3c})\Big|_{(\ln y_2)^2}=0\,,\;(A_{4.4}+A_{3.1})\Big|_{(\ln y_2)^2}=0\,.
\end{equation}

The following real part of the octet (tree) color structure gives almost full
contribution to the correction to the Regge vertex $\gamma_{R_1Q_2}^{Q}$.
\begin{equation}
\begin{split}
&\Re(A_{4.3}+A_{4.3c}+A_{4.4}+A_{3.1}+A_{3.1c}+A_{3.2}+A_{3.2c}+A_{5}+A_{4.1}+A_{4.2})\Big|_{8}=\\
&=A^{Born}\Biggl(\omega_g(q_1)\biggl[\ln y_1+\frac{1}{2}\ln\frac{k_{\bot}^2}{q_{1\bot}^2}\biggr]+\omega_q(q_2)\biggl[\ln y_2+\frac{1}{2}\ln\frac{q_{2\bot}^2}{k_{\bot}^2}\biggr]+\delta_{1Q}^{v,c}+\delta_{1\gamma}^{A}+\delta_{1\gamma}^{c}+\delta_S^c+\delta_S^A\Biggr)+\\
&+A_8^{e}\delta_{2Q}+A_8^{q_1}\delta_{2\gamma}^c-A_8^{q_2}\delta_{3\gamma}.
\label{big sum}
\end{split}
\end{equation}
Together with the mass operator contribution (\ref{self-energy contrib-1}) and
(\ref{self-energy contrib-2}) one obtain the full result for the vertex in question.
In the sum (\ref{big sum}) there are some contributions to the Regge vertices in the
fragmentation region, i.e. $\Gamma_{S'S}^{R_1}$ ~--- see (\ref{GammaSS}) and
$\Gamma_{\gamma' Q}^{Q_2}$~--- see (\ref{GammagammaQ}), and the contributions to the
Regge trajectories.

The vertex correction to the photon production vertex comes from the following
diagram group and reads:
\begin{equation}
\Re(A_{3.5}+A_{3.5c})\Big|_{ (\ln
y_2)^0}=A^{Born}\delta_{1\gamma}^{v}+A^{q_1}_8\delta_{2\gamma}^v+A^{q_2}_8\delta_{3\gamma}.
\label{sum of gamma correction}
\end{equation}
It is obvious from expressions (\ref{big sum}) and (\ref{sum of gamma correction})
that the structure of $A_8^{q_2}$ cancels in the final expression for the Regge
amplitude.

\subsubsection*{The cross-box color structure and the imaginary part of the
amplitude} The cross-box color structure presents only in the  diagrams: $5X$,
$4.1X$, $4.2X$. We introduce the notation for the basic structure, which is
contained in the cross-box color structure:
\begin{equation}
A_{c-b}=-2y_1g^2eT^a_{S'c}T^b_{cS }
\frac{\bar{u}(k_3)t^bt^a\not{q}_{1\bot}
\not{q}_{2\bot}\not{e}_{\bot}u(k_5)}{q_{1\bot}^2q_{2\bot}^2}.
\end{equation}
The cross-box color structure is derived from the amplitude resulting in:
\begin{equation}
\begin{split}
&{\cal
A}\Big|_{crossbox}=g^2\,(-ia_{\Gamma})A_{c-b}(-i\pi)\frac{4}{\epsilon}(-q_{1\bot}^2)^{\epsilon}.
\end{split}
\end{equation}

The imaginary part of  the amplitude contains tree and cross-box color structures:
\begin{equation}
\Im {\cal A}=\pi \frac{(-ia_{\Gamma})g^2}{\epsilon}\biggl[N_cA^{Born}\bigl(
(-q_{1\bot}^2)^{\epsilon}+(-k_{\bot}^2)^{\epsilon}
\bigr)-A_{c-b}4(-q_{1\bot}^2)^{\epsilon}\biggr].
\end{equation}
It is easy to see that the imaginary part does not contain large logarithms ($\ln y_1$ and $\ln y_2$) at all
as it must be according to the Reggeization hypothesis.

\subsubsection*{The final result for the amplitude and the vertex $\gamma_{R_1Q_2}^{Q}$}

Now we present the resulting expression for one-loop corrections to $SQ\rightarrow S'Q'\gamma$ amplitude in the MRK with ${\cal O}(\epsilon)$ accuarcy:
\begin{equation}
\begin{split}
&i{\cal A} =a_{\Gamma}g^2A^{Born}\Biggl\{C_F(-q_{2\bot}^2)^{\epsilon}\Biggl(-\frac{2}{\epsilon^2}+\frac{3}{\epsilon}-9-\frac{2}{\epsilon}\ln y_2+\frac{2\pi^2}{3}+\frac{3q_{1\bot}^2}{q_{1\bot}^2-q_{2\bot}^2}\ln\biggl[\frac{q_{1\bot}^2}{q_{2\bot}^2}\biggr]+\\
&+2\Li_2\biggl[1-\frac{q_{1\bot}^2}{q_{2\bot}^2}\biggr]\Biggr)+N_c(-q_{1\bot}^2)^{\epsilon}\Biggl(-\frac{2}{\epsilon^2}+\frac{1}{\epsilon}\biggl(\frac{1}{3}+\frac{4}{9}\epsilon+n_s\Bigl(\frac{1}{6}-\frac{4}{9}\epsilon\Bigr)+\frac{n_f}{N_c}\Bigl(\frac{2}{3}-\frac{10}{9}\epsilon\Bigr)+\\
&+\ln\biggl[\frac{q_{1\bot}^2}{k_{\bot}^2}\biggr]-2\ln y_1\biggr)-\frac{1}{2}\ln^2\biggl[\frac{q_{1\bot}^2}{k_{\bot}^2}\biggr]+\frac{2\pi^2}{3}-\ln\biggl[\frac{q_{1\bot}^2}{q_{2\bot}^2}\biggr]\ln\biggl[\frac{k_{\bot}^2}{q_{2\bot}^2}\biggr]-2\Li_2\biggl[1-\frac{q_{1\bot}^2}{q_{2\bot}^2}\biggr]\Biggr)\Biggr\}+\\
&+2C_Fa_{\Gamma}g^2A_8^{q_1}+a_{\Gamma}g^2A_8^e\Biggl\{(N_c-C_F)\biggl(1+\frac{k_{\bot}^2}{q_{1\bot}^2-q_{2\bot}^2}-\frac{q_{1\bot}^2k_{\bot}^2}{\bigl(q_{1\bot}^2-q_{2\bot}^2\bigr)^2}\ln\biggl[\frac{q_{1\bot}^2}{q_{2\bot}^2}\biggr]\biggr)+\\
&+C_F\frac{3\,q_{1\bot}^2}{q_{1\bot}^2-q_{2\bot}^2}\ln\biggl[\frac{q_{1\bot}^2}{q_{2\bot}^2}\biggr]\Biggl\}+i\pi \frac{a_{\Gamma}g^2}{\epsilon}\Biggl\{N_cA^{Born}\bigl((-q_{1\bot}^2)^{\epsilon}+ (-k_{\bot}^2)^{\epsilon}\bigr)-A_{c-b}4(-q_{1\bot}^2)^{\epsilon}\Biggr\}.
\label{fullamplitude}
\end{split}
\end{equation}
It is easy to see that the coefficients before structures $A_8^{q_1}$, $A_8^{e}$ are finite in the limit $D\rightarrow 4$.

Comparing  the result (\ref{fullamplitude}) of the amplitude calculation with the expression (\ref{multiReggeform}) for the real part coming from the Reggeization hypothesis we can present  the effective Regge vertex of the quark
production in the central rapidity region in the NLO:
\begin{equation*}
\gamma_{R_1Q_2}^{Q}=-g\frac{1}{k_3^+}\bar{u}(k_3)t^{R_1}\bigl(
\not{q}_{1\perp}+\not{q}_{1\bot}\delta_{1Q}+\not{q}_{2\bot}\delta_{2Q}\bigr),
\end{equation*}
where
\begin{equation*}
\delta_{1Q}=\delta_{1Q}^{s.e.1}+\delta_{1Q}^{s.e.2} +\delta_{1Q}^{v,c}.
\end{equation*}
Contribution from mass operator has the form:
\begin{equation}
\begin{split}
&\delta_{1Q}^{s.e.1}+\delta_{1Q}^{s.e.2} =\frac{N_cg^2(-ia_{\Gamma})}{2}\Biggl\{(-q_{1\bot}^2)^{\epsilon}\Biggl[\frac{n_s}{2}\biggl(\frac{1}{3\epsilon}-\frac{8}{9}\biggr)-\biggl(\frac{5}{3\epsilon}-\frac{31}{9}\biggr)+\frac{n_f}{N_c}\biggl(\frac{4}{3\epsilon}-\frac{20}{9}\biggr)\Biggr]+\\
&+(-q_{2\bot}^2)^{\epsilon}\frac{C_F}{N_c}\biggl(\frac{1}{\epsilon}-1\biggr)\Biggr\}.
\end{split}
\end{equation}
The vertex correction acquires the form with ${\cal O}(\epsilon)$ accuracy:
\begin{equation}
\begin{split}
&\delta_{1Q}^{v,c}=g^2(-ia_{\Gamma})\Biggl\{C_F\Biggl(-\frac{(-k_{\bot}^2)^{\epsilon}}{\epsilon^2}+\frac{(-q_{2\bot}^2)^{\epsilon}}{\epsilon}+\frac{\pi^2}{6}-4+\frac{3q_{1\bot}^2}{q_{1\bot}^2-q_{2\bot}^2}\ln\biggl[\frac{q_{1\bot}^2}{q_{2\bot}^2}\biggr]\Biggr)+\\
&\!\!\!\!\!\!+N_c\Biggl(\frac{(-q_{1\bot}^2)^{\epsilon}}{2\epsilon}+\frac{\pi^2}{6}+3-\frac{1}{2}\ln^2\biggl[\frac{q_{1\bot}^2}{q_{2\bot}^2}\biggr]\Biggr)+\bigl(C_F-N_c\bigr)\Biggl(\frac{1}{2}\ln^2\biggl[\frac{q_{2\bot}^2}{k_{\bot}^2}\biggr]+2\Li_2\biggl[1-
\frac{q_{1\bot}^2}{q_{2\bot}^2}\biggr]\Biggr)\Biggr\}.
\end{split}
\end{equation}
The correction to the term violating helicity has the form that is finite in the limit
$\epsilon\rightarrow0$:
\begin{equation}
\begin{split}
\delta_{2Q}=g^2(-ia_{\Gamma})\Biggl\{&(N_c-C_F)\biggl(1+\frac{k_{\bot}^2}{q_{1\bot}^2-q_{2\bot}^2}-\frac{q_{1\bot}^2k_{\bot}^2}{\bigl(q_{1\bot}^2-q_{2\bot}^2\bigr)^2}\ln\biggl[\frac{q_{1\bot}^2}{q_{2\bot}^2}\biggr]\biggr)+\\
&+C_F\frac{3\,q_{1\bot}^2}{q_{1\bot}^2-q_{2\bot}^2}\ln\biggl[\frac{q_{1\bot}^2}
{q_{2\bot}^2}\biggr]\Biggl\}+{\cal O}(\epsilon).
\end{split}
\end{equation}

For ${\cal N}=4$ SYM case (all particles of the theory are in the adjoint color representation) one can obtain very simple result for the vertex:
\begin{equation}
\begin{split}
&\gamma_{R_1Q_2}^{Q(SYM)}=-\frac{g}{k_3^+}\bar{u}(k_3)T^{R_1}_{QQ_2}\Biggl\{\not{q}_{1\bot}+g^2N_c(-ia_{\Gamma})\Biggl(\not{q}_{1\bot}\biggl[\frac{3}{2\epsilon}\biggl((-q_{1\bot}^2)^{\epsilon}+(-q_{2\bot}^2)^{\epsilon}\biggr)-\\
&-\frac{(-k_{\bot}^2)^{\epsilon}}{\epsilon^2}-\frac{7}{2}+\frac{\pi^2}{3}-\frac{1}{2}\ln^2\frac{q_{1\bot}^2}{q_{2\bot}^2}\biggr]+\bigl(\not{q}_{1\bot}+\not{q}_{2\bot}\bigr)\frac{3q_{1\bot}^2}{q_{1\bot}^2-q_{2\bot}^2}\ln\frac{q_{1\bot}^2}{q_{2\bot}^2}\Biggr)\Biggr\}.
\end{split}
\label{SYMvertex}
\end{equation} 
Here we have used the following substitutions: $C_F\rightarrow N_c$, $\frac{n_f}{N_c}\rightarrow4,\,n_s\rightarrow6-2\epsilon$ (in the dimensional reduction scheme).

\section{Conclusion}
Our paper is devoted to the effective vertex $\gamma_{R_1Q_2}^{Q}$ calculation in the next-to-leading order. The vertex of the massless quark $Q$ production in Reggeon (quark $Q_2$ and gluon $R_1$) collision in t-channels was the last unknown NLO vertex in the central rapidity region to formulate the quark Reggeization hypothesis within the next-to-leading logarithmic approximation. Now all components are ready to perform the hypothesis proof by the bootstrap approach \cite{Fadin:2006bj} having been used in the gluon Reggeization proof both in QCD \cite{FKR:2010-1,FKR:2010-2,FKR:2010-3} and SYM \cite{FKR:2013}. The simplicity of the SYM vertex (\ref{SYMvertex}) gives us an additional tool for the SYM property investigations by use of multi-Regge amplitude form as it was the case for the gluon Reggeization and BDS anzats \cite{Bern:2005iz} in SYM.

In principle, there are some different methods of the effective vertex calculation with t-channel unitarity method being the most popular among them. However in our work we use the straightforward method of one-loop amplitude calculation having equipped with the computer algebra automatization methods based on the Mathematica system and LiteRed package (by R.~N. Lee) \cite{LiteRed} for it. This package is used to reduce integrals emerging in the one-loop amplitude to the basis of master-integrals. Master-integrals in our problem are of known (massless) pentagon and box  types \cite{BDK}. Our method allows us to perform the cross-check by obtaining another elements of the Regge amplitude (quark and gluon trajectories and known effective vertices) as a by-effect.

\subsection*{Acknowledgments}

Work is supported by the Russian Scientific Foundation (grant RFBR 13-02-01023, 15-02-07893).
A.~V. thanks the Dynasty foundation for the financial support.
We would like to thank R.~N.~Lee for helpful comments and discussions.

\section{Appendix}

%
%
%

\subsection{Integrals calculation. Notation}
We reduce all expressions for diagrams \ref{fig:2}  to scalar products and basic
helicity structures. There are two vectors that are not included in the denominators
of the pentagon diagram type in the amplitude. It means that there will be tensor
integrals with up to two indices. There is only one topology of the integral in our
problem:
\begin{equation}
\begin{split}
J_{12345}(n_1,n_2,n_3,n_4,n_5)=J_{12345}(\vec{n})=\int\frac{d^{D}l}
{(2\pi)^{D}}\frac{1}{D_1^{n_1}D_2^{n_2}D_3^{n_3}D_4^{n_4}D_5^{n_5}}\,,
\end{split}
\end{equation}
\[
D_1=l^2\,,\;D_2=(l+k_1)^2\,,\;D_3=(l+k_1+k_2)^2\,,\;D_4=(l+k_1+k_2+k_3)^2\,,\;D_5=(l+k_1+k_2+k_3+k_4)^2
\]
There are only ten master integrals in this topology that have the form
$J_{12345}(\vec{n}_i)\,,\;i=1,\dots,10$, where $\vec{n}_i $ are of the form:
\[
\begin{split}
&\vec{n}_1=(1,1,1,1,1)\,,\;\vec{n}_2=(1,1,1,1,0)\,,\;\vec{n}_3=(1,1,1,0,1)\,,\;\vec{n}_4=(1,1,0,1,1)\,,\;\vec{n}_5=(1,0,1,1,1)\,,\;\\
&\vec{n}_6=(1,0,1,0,0)\,,\;\vec{n}_7=(1,0,0,1,0)\,,\;\vec{n}_8=(0,1,0,1,0)\,,\;\vec{n}_9=(0,1,0,0,1)\,,\;\vec{n}_{10}=(0,0,1,0,1)\,,\;
\end{split}
\]
and one has four integral types with the same topology  with them having different
permutations of the external momenta:
\[
\begin{array}{|c|c|c|c|c|c|}
\hline & D_1 & D_2 & D_3 & D_4 & D_5 \\ \hline J_{12345} & l^2 & (l+k_1)^2 &
(l+k_1+k_2)^2 & (l+k_1+k_2+k_3)^2 & (l+k_1+k_2+k_3+k_4)^2\\ \hline J_{21345} & l^2 &
(l+k_2)^2 & (l+k_1+k_2)^2 & (l+k_1+k_2+k_3)^2 & (l+k_1+k_2+k_3+k_4)^2\\ \hline
J_{12435} & l^2 & (l+k_1)^2 & (l+k_1+k_2)^2 & (l+k_1+k_2+k_4)^2 &
(l+k_1+k_2+k_3+k_4)^2\\ \hline J_{41235} & l^2 & (l+k_4)^2 & (l+k_1+k_4)^2 &
(l+k_1+k_2+k_4)^2 & (l+k_1+k_2+k_3+k_4)^2\\ \hline
\end{array}
\]
There are only three types of different master integrals.  The first one is a
pentagon with massless external lines (for example, $J_{12345}(\vec{n}_1)$). The
second one is a box with one external line with mass (for example,
$J_{12345}(\vec{n}_2)$). And the third type is a bubble (for example,
$J_{12345}(\vec{n}_6)$).

The following table shows the integrals used in the diagrams in Fig.~\ref{fig:2}:
\[
\begin{array}{|c|c|c|c|c|}
\hline $Integrals$ & $Diagrams$ \\ \hline
J_{12345} & d5,\,d4.1,\,d4.2,\,d4.4,\,d3.X,\,d2.X \\
J_{21345} & d5c,\,d4.1c,\,d4.2c \\
J_{12435} & d4.3\\
J_{41235} & d4.3c \\ \hline
\end{array}
\]

\subsection{Tensor momentum integrals}

We introduce the following notation $I[.]$ for various integrands containing the
argument of the square bracket. For instance,
\begin{equation}
\begin{split}
&I\bigl[l^{\mu}\bigr]\equiv\int\frac{d^{D}l}{(2\pi)^{D}}\frac{l^{\mu}}{D_1^{n_1}D_2^{n_2}D_3^{n_3}D_4^{n_4}D_5^{n_5}}
\end{split}
\end{equation}
Integral with an external index $\mu$ is expressed as a linear  combination of the
incoming momenta:
\begin{equation}
I\bigl[l^{\mu}\bigr]=\sum_ik_i^{\mu}I[a_i],
\end{equation}
where $a_i$ are scalar polynomial functions of $l^{\mu}$. It is easy to express the
integral with an external index as a linear combination of integrals without
external indices. Since we have four independent vectors $k_i$ (that are in the four-dimensional
subspace), so the matrix $ m_ {ij} $ is invertible. It is easy to find that
eventually
\begin{equation}
I\bigl[l^{\mu}\bigr]=\sum_{ij}k_i^{\mu}m^{-1}_{ij}I\bigl[(k_j,l)\bigr]\,,\;(k_i,k_j)=m_{ij}\,,
\end{equation}
where $I\bigl[(k_j,l)\bigr] $ is expressed in terms  of integrals with other powers
of denominators.

Tensor integral with two indices is expressed through the metric tensor subspace $
D-4 $, and entering into integral momenta
\begin{equation}
I\bigl[l^{\mu}l^{\nu}\bigr]=g^{\mu\nu}_{D-4}I[A]+
\sum_{i,j,r,n}m^{-1}_{ij}m^{-1}_{rn}\,k_i^{\mu}k_r^{\nu}\,I\bigl[(k_j,l)(k_n,l)\bigr],
\end{equation}
\[g^{\mu\nu}_{D-4}k_{i\mu}=0\,,\;g^{\mu\nu}_{D-4}g_{\mu\nu}=D-4.\]

The coefficient before the metric tensor can be easily calculated:
\begin{equation}
I\bigl[A\bigr]=\frac{1}{D-4}\Bigl(
I\bigl[l^2\bigr]-\sum_{ij}m^{-1}_{ij}I\bigl[(l,k_i)(l,k_j)\bigr]\Bigr),
\end{equation}
\begin{equation}
\begin{split}
&I[l^{\mu}l^{\nu}]=\frac{g^{\mu\nu}-m^{-1}_{ij}k_i^{\mu}k_j^{\nu}}{D-4}\biggl(I[l^2]-m^{-1}_{rn}I[(l,k_r)(l,k_n)]\biggr)+m^{-1}_{ij}m^{-1}_{rn}\,k_i^{\mu}k_r^{\nu}I[(k_j,l)(k_n,l)]
\end{split}
\end{equation}

The integrals with three external Lorentz indices will not arise in our problem. A
loop momentum convoluted with the momentum included in the denominator is easily
expressed in terms of a linear combination of the denominators. Integrals with two
indices appear only in the expression of $I[(e,l)\not{l}]$.

\subsection{Master Integrals}
We work in the dimensional regularization with $D=4+2\epsilon$. Hereafter we use the
notation for the common multiplier emerging in integral calculations.
\begin{equation}\label{eq:aG}
a_{\Gamma}=i\frac{\Gamma\bigl(3 - \frac{D}{2}\bigr)\Gamma^2\bigl(\frac{D}{2} -
1\bigl)}{(4\pi)^{\frac{D}{2}}\Gamma\bigl(D-3\bigr)}.
\end{equation}

In the master integral expressions \cite{BDK} we assume all the invariants to be
negative. To continue analytically these expressions to the physical region of our
process (See Fig~\ref{fig:1}) it is necessary to make a prescription
$(k_i+k_j)^2=s_{ij}\rightarrow s_{ij}+i0$ for the invariants involved.

There are three principal master integrals in our calculation \cite{BDK}:
\[
J_{12345}(1,0,1,0,0)=\int\frac{d^{D}l}{(2\pi)^D}
\frac{1}{l^2(l+k_1+k_2)^2}=-a_{\Gamma}\frac{(-t_1)^{\epsilon}}{\epsilon(1+2\epsilon)}\,,\;t_1<0;
\]
\[
\begin{split}
&J_{12345}(1,1,1,1,0)=\int\frac{d^{D}l}{(2\pi)^{D}}\frac{1}{l^2(l+k_1)^2
(l+k_1+k_2)^2(l+k_1+k_2+k_3)^2}=\\
&=2a_{\Gamma}\frac{(-t_2)^{-\epsilon}}{(-t_1)^{1-\epsilon}(-s_1)^{1-\epsilon}}\biggl[
\frac{1}{\epsilon^2}+\Li_2\Bigl(1-\frac{t_1}{t_2}\Bigr)+
\Li_2\Bigl(1-\frac{s_1}{t_2}\Bigr)-\frac{\pi^2}{6}\biggr]+{\cal O}(\epsilon),\\
&s_1<0,\;t_1<0,\;t_2<0;
\end{split}
\]
\[
\begin{split}
&J_{12345}(1,1,1,1,1) =\int\frac{d^{D}l}{(2\pi)^{D}}\frac{1}{l^2(l+k_1)^2(l+k_1+k_2)^2(l+k_1+k_2+k_3)^2(l+k_1+k_2+k_3+k_4)^2}=\\
&=-a_{\Gamma}\Biggl\{
\frac{(-s)^{-\epsilon}(-t_1)^{-\epsilon}}{(-s_1)^{1-\epsilon}(-s_2)^{1-\epsilon}(-t_2)^{1-\epsilon}}\biggl(\frac{1}{\epsilon^2}+2\Li_2\Bigl(1-\frac{s_1}{s}\Bigr)+2\Li_2\Bigl(1-\frac{t_2}{t_1}\Bigr)-\frac{\pi^2}{6}\biggr)+\\
&+\frac{(-t_1)^{-\epsilon}(-s_1)^{-\epsilon}}{(-s_2)^{1-\epsilon}(-t_2)^{1-\epsilon}(-s)^{1-\epsilon}}\biggl(\frac{1}{\epsilon^2}+2\Li_2\Bigl(1-\frac{s_2}{t_1}\Bigr)+2\Li_2\Bigl(1-\frac{s}{s_1}\Bigr)-\frac{\pi^2}{6}\biggr)+\\
&+\frac{(-s_1)^{-\epsilon}(-s_2)^{-\epsilon}}{(-t_2)^{1-\epsilon}(-s)^{1-\epsilon}(-t_1)^{1-\epsilon}}\biggl(\frac{1}{\epsilon^2}+2\Li_2\Bigl(1-\frac{t_2}{s_1}\Bigr)+2\Li_2\Bigl(1-\frac{t_1}{s_2}\Bigr)-\frac{\pi^2}{6}\biggr)+\\
&+\frac{(-s_2)^{-\epsilon}(-t_2)^{-\epsilon}}{(-s)^{1-\epsilon}(-t_1)^{1-\epsilon}(-s_1)^{1-\epsilon}}\biggl(\frac{1}{\epsilon^2}+2\Li_2\Bigl(1-\frac{s}{s_2}\Bigr)+2\Li_2\Bigl(1-\frac{s_1}{t_2}\Bigr)-\frac{\pi^2}{6}\biggr)+\\
&+\frac{(-t_2)^{-\epsilon}(-s)^{-\epsilon}}{(-t_1)^{1-\epsilon}(-s_1)^{1-\epsilon}(-s_2)^{1-\epsilon}}\biggl(\frac{1}{\epsilon^2}+2\Li_2\Bigl(1-\frac{t_1}{t_2}\Bigr)+2\Li_2\Bigl(1-\frac{s_2}{s}\Bigr)-\frac{\pi^2}{6}\biggr)\Biggr\}
+{\cal O}(\epsilon),\\
&s<0,\;s_1<0,\;s_2<0,\;t_1<0,\;t_2<0;
\end{split}
\]
Here we use notations (\ref{invariants}) for kinematic invariants.

In the analytical continuation to the physical domain in polylogarithmic functions
one needs to choose the correct branch using the following properties:
\[
\Li_2(x\pm i0)=\frac{\pi^2}{3}-\frac{1}{2}\ln^2x-\Li_2(x^{-1})\pm i\pi\ln x\;,\quad
x>1
\]
Providing $\sigma_1,\;\sigma_2$ to be the signs of $s_1,\;s_2$,  for the case
$\sigma_1\sigma_2=-1$ one has
\[
\Li_2\Bigl(1-\frac{s_1}{s_2}\Bigr)=\frac{\pi^2}{3}-
\frac{1}{2}\ln^2\Bigl(1-\frac{s_1}{s_2}\Bigr)-
\Li_2\Bigl(\frac{1}{1-\frac{s_1}{s_2}}\Bigr)+i\sigma_1\pi\ln\Bigl(1-\frac{s_1}{s_2}\Bigr).
\]



\begin{thebibliography}{25}

\bibitem{Balitsky:1978ic}
I.~I.~Balitsky and L.~N.~Lipatov,
Sov.\ J.\ Nucl.\ Phys.\  {\bf 28} (1978) 822 [Yad.\ Fiz.\  {\bf 28} (1978) 1597].

\bibitem{Fadin:1975cb}
V.~S.~Fadin, E.~A.~Kuraev and L.~N.~Lipatov,
Phys.\ Lett.\  B {\bf 60} (1975) 50.

\bibitem{Kuraev:1976ge}
E.~A.~Kuraev, L.~N.~Lipatov and V.~S.~Fadin,
  Zh. Eksp. Teor. Fiz. \textbf{71} (1976) 840 [Sov. Phys. JETP \textbf{44}
(1976) 443].

\bibitem{Kuraev:1977fs}
E.~A.~Kuraev, L.~N.~Lipatov and V.~S.~Fadin,
Zh.\ Eksp.\ Teor.\ Fiz.\ {\bf 72} (1977) 377 [Sov.\ Phys.\ JETP {\bf 45} (1977)
199].


\bibitem{Lipatov:1976zz}
L.~N.~Lipatov,
Sov.\ J.\ Nucl.\ Phys.\  {\bf 23}, 338 (1976).

\bibitem{BG:2007} A.~V. Bogdan, A.~V. Grabovsky,
Nucl. Phys. B773 (2007) 65--83.

\bibitem{Fadin:2006bj}
V.~S.~Fadin, R.~Fiore, M.~G.~Kozlov, and A.~V.~Reznichenko,
Phys.\ Lett.\  B {\bf 639} (2006) 74 [arXiv:hep-ph/0602006].

\bibitem{BF:2006} A.~V. Bogdan and V.~S. Fadin,
Nucl. Phys. B 740 (2006) 36--57.

\bibitem{Fadin:2003xs}
V.~S.~Fadin, M.~G.~Kozlov and A.~V.~Reznichenko,
Yad.\ Fiz.\  {\bf 67}, 377 (2004) [Phys.\ Atom.\ Nucl.\  {\bf 67}, 359 (2004)],
[hep-ph/0302224].

\bibitem{FKR:2010-1}
V.~S. Fadin, M.~G. Kozlov, A.~V. Reznichenko,
Phys.Atom.Nucl. vol. \textbf{74}, (2011) pp. 758-770 [Preprint INP 2010-26].

\bibitem{FKR:2010-2}
V.~S.~Fadin, M.~G.~Kozlov, and A.~V.~Reznichenko,
Phys.Atom.Nucl. vol. \textbf{75}, (2012), pp. 850-865 [Preprint INP 2011-23].


\bibitem{FKR:2010-3}
V.~S.~Fadin, M.~G.~Kozlov, and A.~V.~Reznichenko,
Phys.Atom.Nucl., vol. \textbf{75},  (2012) pp. 493-506  [Preprint INP 2011-24].

\bibitem{FKR:2013}
V.~S. Fadin,  M.~G. Kozlov, A.~V. Reznichenko,
Phys.Atom.Nucl., vol. \textbf{77} (2014), pp.  251-273 [Preprint INP 2012-32].

\bibitem{BDK}
Z.~Bern, L.~Dixon, and D.~Kosower,
Nucl.Phys. B \textbf{412} (1994) pp. 751-816 [SLAC-PUB-5947, hep-ph/9306240].

\bibitem{LiteRed}
R.~N.~Lee,  arXiv:1212.2685 (2012).


\bibitem{FF:2001}
V. S. Fadin, R. Fiore, Phys.Rev. D64 (2001) 114012

\bibitem{LV:2001}
L. N. Lipatov and M. I. Vyazovsky, Nucl. Phys. B 597 (2001), 399, arXiv:hep-ph/0009340


\bibitem{F:2003}
V. S. Fadin, Phys. Atom. Nucl. 66, 2017 (2003).

\bibitem{FG:2008}
R.~E.~Gerasimov, V.~S.~Fadin,
Phys.Atom. Nucl. 73 (2010) 1214--1228, Preprint INP 2008-36.

\bibitem{BG:2006}
A.V. Bogdan, A.V. Grabovsky,Nucl.Phys. B 757 (2006), 211-232

\bibitem{FL:1993}
V. S. Fadin and  L.N. Lipatov, Nucl. Phys. B 406 (1993), 259--292

\bibitem{FFP:2001}
V. S. Fadin, R. Fiore and A. Papa, Phys. Rev. D 63 (2001), 034001, arXiv:hep-ph/008006

\bibitem{DS:1999}
V. Del Duca and C. R. Schmidt, Phys. Rev. D 59 (1999), 074004



\bibitem{Bern:2005iz} 
Z.~Bern, L.~J.~Dixon and V.~A.~Smirnov,
Phys.\ Rev.\ D {\bf 72} (2005) 085001 [hep-th/0505205].

\end{thebibliography}
\end{document}